\documentclass[10pt,prc,amsmath,amssymb,twocolumn,aps,showpacs,superscriptaddress,groupedaddress]{revtex4-1}
\usepackage{amsmath}
\usepackage{graphicx}
\usepackage[dvips]{color}
\newcommand{\n}[1]{\ensuremath{|\mathbf{#1}|}}
\newcommand{\ve}[1]{\ensuremath{\mathbf{#1}}}

\newcommand{\be}{\begin{equation}}
\newcommand{\ee}{\end{equation}}
\newcommand{\ba}{\begin{eqnarray}}
\newcommand{\ea}{\end{eqnarray}}

\begin{document}

\title{Charged-current quasielastic neutrino cross sections on $^{12}$C with realistic spectral and scaling functions}

\author{M.V.~Ivanov}
\affiliation{Grupo\,de\,F\'{i}sica\,Nuclear,\,Departamento\,de\,F\'{i}sica\,At\'omica,\,Molecular\,y\,Nuclear, Facultad\,de\,Ciencias\,\,F\'{i}sicas,\,Universidad\,Complutense\,de\,Madrid,\,Madrid\,E-28040,\,Spain}
\affiliation{Institute\,for\,Nuclear\,Research\,and\,Nuclear\,Energy,\,Bulgarian\,Academy\,of\,Sciences,\,Sofia\,1784,\,Bulgaria}

\author{A.~N.~Antonov}
\affiliation{Institute\,for\,Nuclear\,Research\,and\,Nuclear\,Energy,\,Bulgarian\,Academy\,of\,Sciences,\,Sofia\,1784,\,Bulgaria}

\author{J.A. Caballero}
\affiliation{Departamento de F\'{\i}sica At\'{o}mica, Molecular y Nuclear, Universidad de Sevilla, 41080 Sevilla, Spain}

\author{G.D. Megias}
\affiliation{Departamento de F\'{\i}sica At\'{o}mica, Molecular y Nuclear, Universidad de Sevilla, 41080 Sevilla, Spain}

\author{M.B. Barbaro}
\affiliation{Dipartimento di Fisica, Universit\`a di Torino and  INFN, Sezione di Torino, Via P. Giuria 1, 10125 Torino, Italy}

\author{E.~Moya de Guerra}
\affiliation{Grupo\,de\,F\'{i}sica\,Nuclear,\,Departamento\,de\,F\'{i}sica\,At\'omica,\,Molecular\,y\,Nuclear, Facultad\,de\,Ciencias\,\,F\'{i}sicas,\,Universidad\,Complutense\,de\,Madrid,\,Madrid\,E-28040,\,Spain}

\author{J.M. Ud\'{\i}as}
\affiliation{Grupo\,de\,F\'{i}sica\,Nuclear,\,Departamento\,de\,F\'{i}sica\,At\'omica,\,Molecular\,y\,Nuclear, Facultad\,de\,Ciencias\,\,F\'{i}sicas,\,Universidad\,Complutense\,de\,Madrid,\,Madrid\,E-28040,\,Spain}

\begin{abstract}

Charge-current quasielastic (CCQE) (anti)neutrino scattering cross sections on a $^{12}$C target are analyzed using a spectral function $S(p,{\cal E})$ that gives a scaling function in accordance with the ($e,e'$) scattering data. The spectral function accounts for the nucleon-nucleon (\emph{NN}) correlations, it has a realistic energy dependence and natural orbitals (NO's) from the Jastrow correlation method are used in its construction. In all calculations the standard value of the axial mass $M_A= 1.032$~GeV/c$^2$ is used. The results are compared with those when \emph{NN} correlations are not included, as in the Relativistic Fermi Gas (RFG) model, or when harmonic-oscillator (HO) single-particle wave functions are used instead of NO's. The role of the final-state interactions (FSI) on the theoretical spectral and scaling functions, as well as on the cross sections is accounted for. A comparison of the results for the cases with and without FSI, as well as to results from the phenomenological scaling function obtained from the superscaling analysis (SuSA) is carried out. Our calculations based on the impulse approximation (IA) underpredict the MiniBooNE data, but agree with the data from the NOMAD experiment. The possible missing ingredients in the considered theoretical models are discussed.

\end{abstract}

\pacs{25.30.Pt, 13.15.+g, 24.10.Jv}

\maketitle

\section{Introduction \label{sec:1}}

The recent MiniBooNE data on charged-current quasielastic (CCQE) scattering of muon neutrino on nuclei~\cite{miniboone, miniboone-ant} have revealed the important role played by the nuclear and nucleonic ingredients necessary for the description of the reaction. Many theoretical works have been devoted to analyses of empirical data (see, \emph{e.g.}~\cite{amaro2013, Amaro2011151, PhysRevD.84.033004, PhysRevLett.108.152501, PhysRevLett.105.132301, PhysRevC.81.045502, PhysRevC.87.065501, PhysRevC.82.045502, PhysRevC.82.055501, Meucci:2011nc, Meucci:2011vd, bodek_neutrino_2011, Nieves201272, Nieves201390, PhysRevC.86.014614, PhysRevC.86.015504, barbaro2011, Pandey2013, Ankowski:2012ei}). It turned out, unexpectedly, that the cross  sections are underestimated by traditional nuclear models, such as the Relativistic Fermi Gas (RFG) model, the RPA calculations, the relativistic Greens's function approaches, the Relativistic Mean Field (RMF) theory, the use of realistic spectral functions and others. It was shown, however, that the accordance between theory and data can be recovered within the simple RFG, if an unusually large value of the axial mass $M_A \cong 1.35$~GeV/c$^2$ (as compared to the standard value $M_A= 1.032$~GeV/c$^2$) is employed in the dipole parametrization of the nuclear axial form factor. At the same time, the necessity to account for the multi-nucleon excitations (in particular, two particle emission) has been proposed in \emph{e.g.}~\cite{Amaro2011151, PhysRevD.84.033004, PhysRevC.81.045502, PhysRevC.87.065501, Nieves201272} and a good agreement with the MiniBoonE data has been shown in Ref.~\cite{PhysRevC.81.045502, PhysRevC.87.065501, Nieves201272} using the standard $M_A$-value. The calculations based on the exact relativistic  account for the Meson Exchange Currents (MEC) within the $2p-2h$ RFG approach give an enhancement of the cross sections but do not fully describe the discrepancy between the data and the theory.

It should be pointed out, however, that the data from CCQE $\nu_\mu$($\overline{\nu}_\mu$)-$^{12}$C cross section measurements from $3$ to $100$~GeV of the NOMAD Collaboration~\cite{lyubushkin_study_2009} do not impose an anomalously large axial-vector mass to be used and do not match with the lower-energy MiniBooNE data. It is worth mentioning also that the recent data on CCQE $\nu_\mu$($\overline{\nu}_\mu$)-$^{12}$C from the MINERvA Collaboration~\cite{Minerva2013, Minerva2013a} disfavor the value $M_A \simeq 1.35$~GeV/c$^2$. So, a consistent theoretical analysis of the cross sections in the entire energy range ($0-100$~GeV) is missing. Therefore it is highly desirable to provide a consistent framework that describes successfully the QE electron data from intermediate up to very high energies using a \textbf{relativistic} nuclear model.

The experiments on neutrino-nuclei scattering are of great importance. The neutrino properties, and particularly the parameters of their oscillations, make it possible to obtain information about the limits of the Standard Model. In the mentioned experiments the interaction of the neutrino occurs with nucleons bound in nuclei. The analyses of such processes within different methods involve various effects such as nucleon-nucleon (NN) correlations, the final state interactions (FSI), possible modifications of the nucleon properties inside the nuclear medium and others. These effects, however, cannot be presently accounted for in an unambiguous and precise way, and what is most important, in most cases they are highly model-dependent. A possible way to avoid model-dependencies is to use the nuclear response to other leptonic probes, such as electrons, under similar conditions to the neutrino experiments. The superscaling approximation (SuSA) follows this general trend. The analyses of scaling~\cite{annurev.ns.40.120190.002041, PhysRevLett.45.871, PhysRevC.36.1208, PhysRevC.39.259, PhysRevC.43.1155, PhysRevC.46.1045, PhysRevC.53.1689, degliAtti1999447} and superscaling~\cite{PhysRevLett.82.3212, PhysRevC.60.065502, PhysRevC.65.025502, PhysRevC.74.054603, PhysRevC.38.1801, Barbaro1998137, PhysRevC.69.035502, PhysRevC.69.044321, PhysRevC.71.014317, PhysRevC.73.047302, PhysRevC.77.034612} phenomena observed in electron scattering on nuclei have led to the use of the scaling function directly extracted from ($e,e'$) data to predict (anti)neutrino-nucleus cross sections~\cite{PhysRevC.71.015501}, just avoiding the usage of a particular nuclear structure model. A ``superscaling function'' $f(\psi)$ has been extracted from the data by factoring –out the single-nucleon content of the double-differential cross section and plotting the remaining nuclear response versus a scaling variable $\psi(q, \omega)$ ($q$ and $\omega$ being the momentum transfer and transferred  energy, respectively). For high enough values of the momentum transfer (roughly $q>400$~MeV/c) the explicit dependence of $f(\psi)$ on $q$ is very weak at transferred energies below the quasielastic peak (scaling of the first kind). Scaling of second kind (\emph{i.e.} no dependence of $f(\psi)$ on the mass number $A$) turns out to be excellent in the same region. The term ``superscaling'' means the occurrence of both first and second types of scaling.

The observation  of superscaling in the data on inclusive electron-nucleus scattering~\cite{PhysRevLett.82.3212, PhysRevC.60.065502} has justified the extraction of an universal nuclear response to be applied to processes with weak interacting probes. The RFG model, employed in most analyses of neutrino experiments, does exhibit supescaling~\cite{PhysRevC.38.1801, Barbaro1998137}, but the corresponding nuclear response cannot explain successfully the electron scattering data. This imposes the necessity to use more complex dynamical pictures of the nuclei (beyond the RFG) for the description of the nuclear response at  intermediate energies. The SuSA results for CCQE (anti)neutrino scattering have been based on the empirical superscaling function extracted from the world data on QE electron scattering~\cite{Jourdan1996117}. Later SuSA has been  applied~\cite{PhysRevC.73.035503} to neutral current scattering and extended to the $\Delta$-resonance region~\cite{PhysRevC.71.015501} as well.

The SuSA approach has been already employed to describe the non-pionic (QE) cross section of the MiniBooNE $\nu_\mu$($\overline{\nu}_\mu$) nucleus cross section~\cite{Amaro2011151, PhysRevD.84.033004, PhysRevLett.108.152501} and in Ref.~\cite{amaro2013} it has been applied to (anti)neutrino CCQE on $^{12}$C for energy range up to 100~GeV with a comparison with the MiniBooNE and NOMAD~\cite{lyubushkin_study_2009} data. In Ref.~\cite{Ivanov2012178} SuSA has been used to analyze CC pion ($\pi^+$) production cross section measured at  MiniBooNE~\cite{PhysRevD.83.052007, PhysRevD.83.052009}, incorporating effects of FSI, the properties of the $\Delta$-resonance in the nuclear medium, as well as both the contributions of coherent and incoherent productions.

The investigations of inclusive QE electron-nucleus scattering make it possible to obtain information about the main characteristics of nuclear structure, namely, the spectral function $S(p,{\cal E})$ and the nucleon momentum distribution $n(p)$. This possibility is based on the validity of scaling arguments that has been clearly demonstrated in the analyses of world ($e,e'$) data revealing also the specific shape of the scaling function, with a significant tail to high positive values of the scaling variable. It has been shown in detail in Ref.~\cite{PhysRevC.81.055502} that the important connection between the scaling function [given directly from the ($e,e'$) data analysis] and $S(p,{\cal E})$ or $n(p)$ exists only under very restrictive conditions, namely: (i) the PWIA in the description of the reaction mechanism, and (ii) additional assumptions on the integration limits consistent with the kinematically allowed region.

The area of analyses of the scaling function, the spectral function and their connection (see, \emph{e.g.}~\cite{PhysRevC.81.055502, PhysRevC.83.045504}) provides insight into the validity of the mean-field approximation (MFA) and the role of the \emph{NN} correlations, as well as into the effects of FSI. Though in the MFA it is possible, in principle, to obtain the contributions of different shells to $S(p,{\cal E})$ and $n(p)$ for each single-particle state, due to the residual interactions the hole states are not eigenstates of the residual nucleus but are mixtures of several single-particle states. The latter leads to the spreading of the shell structure and requires studies of the spectral function using theoretical methods going beyond the MFA in order to describe successfully the relevant experiments. In Ref.~\cite{PhysRevC.83.045504} a realistic spectral function $S(p,{\cal E})$ has been constructed being in agreement with the scaling function $f(\psi)$ obtained from the ($e,e'$) data. For this purpose  effects beyond MFA have been considered. The procedure included: (i) the account for effects of a finite energy spread, and (ii) the account for \emph{NN} correlation effects considering single-particle momentum distributions $n_i(p)$ (that are components of $S(p,{\cal E})$) beyond the MFA, such as those related to the usage of natural orbitals (NO's)~\cite{PhysRev.97.1474} for the single-particle wave functions and occupation numbers within methods in which short-range \emph{NN} correlations are included. For the latter the Jastrow correlation method~\cite{PhysRevC.48.74} has been considered. Also in Ref.~\cite{PhysRevC.83.045504} FSI were accounted for using complex optical potential that has given a spectral function $S(p,{\cal E})$ leading to asymmetric scaling function in accordance with the experimental analysis, thus showing the essential role of the FSI in the description of electron scattering reactions.

The aim of the present paper is to continue our work from Ref.~\cite{Ivanov2012178} but using the results obtained in Ref.~\cite{PhysRevC.83.045504} for a realistic spectral function $S(p,{\cal E})$ instead of the phenomenological SuSA approach. The spectral function from our previous work~\cite{PhysRevC.83.045504} will be applied to analysis of CCQE (anti)neutrino cross sections on a $^{12}$C target measured by the MiniBooNE Collaboration~\cite{miniboone, miniboone-ant} for neutrino energies in the $1$~GeV region and also, extending the range of the energy, up to $100$~GeV~\cite{lyubushkin_study_2009}. Our approach includes i) realistic energy dependence of the spectral function $S(p,{\cal E})$ and ii) an account for the effects of short-range \emph{NN} correlations when NO's from the Jastrow method are included. These results will be compared with those when the \emph{NN} correlations are not included. Second, the role of the FSI on the spectral function and cross sections will be shown comparing the corresponding results for the cases RFG+FSI, HO+FSI and NO+FSI with those without accounting for the FSI. We present results for the scaling function $f(\psi)$, for the double-differential cross section $d^2\sigma/dT_\mu d\cos\theta_\mu$, as well as for those when the latter is integrated over the muon scattering angle ($\langle d\sigma/dT_\mu\rangle$), or over the muon kinetic energy ($\langle d\sigma/d\cos\theta_\mu\rangle$) and, finally, for the total cross section of $\nu_\mu$($\overline{\nu}_\mu$)-$^{12}$C scattering.

The theoretical scheme of the work is given in Sec.~\ref{sec:2}. It contains, first, the methods to obtain a realistic spectral function, and second, the main relationships concerning CCQE neutrino-nucleus reaction cross section. The results of the calculations and discussion are presented in Sec.~\ref{sec:3}. A summary of the work and our conclusions are given in Sec.~\ref{sec:4}.

\section{THEORETICAL SCHEME \label{sec:2}}

\subsection{Inclusive electron-nuclei cross section, spectral function and scaling function\label{sec:2.1}}

Within the PWIA (see~\cite{PhysRevC.81.055502, PhysRevC.83.045504} and references therein) the differential cross section for the ($e,e'N$) process factorizes in the form:
\be
\left[\frac{d\sigma}{d\epsilon'd\Omega'dp_Nd\Omega_N}
\right]_{(e,e'N)}^{PWIA}= K\sigma^{eN}(q,\omega;p,{\cal
E},\phi_N)S(p,{\cal E})\,,\label{PWIA}
\ee
where $\sigma^{eN}$ is the electron-nucleon cross section for a moving off-shell nucleon, $K$ is a kinematical factor~\cite{Raskin198978} and $S(p,{\cal E})$ is the spectral function giving the probability to find a nucleon of certain momentum and energy in the nucleus~\cite{Frullani1984, Boffi19931, boffibook, Kelly:1996hd, PhysRevC.51.1800}. In Eq.~(\ref{PWIA}) ${\cal E}$ is the excitation energy that is essentially the missing energy minus the separation energy and $p$ is the missing momentum. If the spectral function is assumed to be isospin independent and $\sigma ^{eN}$ to have a very mild dependence on $p$ and ${\cal E}$, then the scaling function $F(q,\omega )$ can be represented in PWIA as a ratio
\be F(q,\omega)\cong
\dfrac{\left[d\sigma/d\epsilon'd\Omega'\right]_{(e,e')}}{\overline{\sigma}^{eN}
(q,\omega;p=|y|,{\cal E}=0)}\,, \label{scaling}
\ee
between the differential cross section for inclusive QE ($e,e'$) scattering and the azimutal angle-averaged single-nucleon cross section $\overline{\sigma}^{eN}$
\be
\overline{\sigma}^{eN}\equiv
K\sum_{i=1}^A\int d\phi_{N_i}\dfrac{\sigma^{eN_i}}{2\pi}. \label{sn}
\ee
In Eq.~(\ref{scaling}) $\overline{\sigma}^{eN}$ is taken at $p= |y|$, where the magnitude of the scaling variable $y$ is the smallest value of $p$ that can occur in electron-nucleus scattering for the smallest possible value of the excitation energy (${\cal E}=0$), \emph{i.e.} at the smallest value of the missing energy. In PWIA the scaling function~(\ref{scaling}) can be expressed by the spectral function
\be F(q,\omega)=
2\pi\int\!\!\!\int_{\Sigma(q,\omega)}p\,dp\, d{\cal E}\,S(p,{\cal
E}) \, , \label{scaling_function}
\ee
where ${\Sigma(q,\omega)}$ represents the kinematically allowed region~\cite{PhysRevC.81.055502}. Only when the region ${\Sigma(q,\omega)}$ can be extended to infinity in the excitation energy plane ({\it i.e.,} at ${\cal E}_{\max}\rightarrow \infty$), the scaling function may be related to the momentum distribution $n(p)$ of the nucleus:
\be
n(p)= \int_0^\infty d{\cal
E} S(p,{\cal E}).\label{nk}
\ee
It was emphasized in~\cite{PhysRevC.81.055502} that Eq.~(\ref{scaling_function}) cannot be applied to the empirically extracted scaling function $F_\text{exp} (q,\omega )$ (that at high values of the momentum transfer $q$ becomes a function only of a scaling variable $y$ and not of $q$~\cite{PhysRevLett.82.3212, PhysRevC.60.065502, PhysRevC.65.025502}) because of ingredients not included in the PWIA, such as the final state interaction, meson-exchange currents, rescattering processes, {\it etc.}

Using as a guide the RFG model, it became possible to introduce three ``universal'' experimental dimensionless superscaling functions $f^{L(T)}_\text{exp}(q,\omega) \equiv k_A F^{L(T)}_\text{exp}(q,\omega)\,$, where $k_A$ is a phenomenological momentum scale for a specific nucleus (being the Fermi momentum $k_F$ in the case of RFG model). The letters $L$ and $T$ denote the longitudinal and transverse functions, respectively.

In~\cite{PhysRevC.83.045504} it was made an attempt to extract more information about the spectral function $S(p,{\cal E})$ from the experimentally known scaling function (under the restrictions of the PWIA). First, it was constructed within the independent particle shell model (IPSM):
\begin{equation}\label{HF}
    S_{IPSM}(p,{\cal E})=\sum_{i}2(2j_i+1)n_i(p) \delta({\cal E}-{\cal E}_i),
\end{equation}
where $n_i(p)$ is the momentum distribution of the shell-model single-particle state $i$ and ${\cal E}_i$ is the eigenvalue of the $i$-state energy. Second, when effects beyond the mean-field approximation are considered, the energy dependence would be better represented by a function with a finite width in energy instead of by a $\delta$-function. Hence, the latter can be replaced by a Gaussian distribution $G_{\sigma _i}({\cal E} - {\cal E}_i)$:
\begin{equation}\label{HF+Gauss}
    S(p,{\cal E})=\sum_{i}2(2j_i+1)n_i(p) G_{\sigma_i}({\cal E}-{\cal E}_i),
\end{equation}
where
\begin{equation}\label{eq3a}
G_{\sigma_i}({\cal E}-{\cal E}_i)=
\dfrac{1}{\sigma_i\sqrt{\pi}}e^{-\frac{({\cal E}-{\cal E}_i)^2}{\sigma_i^2}}
\end{equation}
$\sigma_i$ being a parameter related to the width of the hole state $i$.

In Ref.~\cite{PhysRevC.83.045504} it was also considered another form of the energy dependence, namely the Lorentzian function $L_{\Gamma_i}({\cal E} - {\cal E}_i)$:
\begin{equation}\label{HF+lorent}
    S(p,{\cal E})=\sum_{i}2(2j_i+1)n_i(p) L_{\Gamma_i}({\cal E} - {\cal E}_i),
\end{equation}
with
\begin{equation}\label{lorent}
L_{\Gamma_i}({\cal E}-{\cal E}_i)=
\dfrac{1}{\pi}\dfrac{\Gamma_i/2}{({\cal E}-{\cal E}_i)^2+(\Gamma_i/2)^2}\, ,
\end{equation}
where ${\Gamma_i}$ is the width for a given single-particle hole state $i$.

Starting with a momentum distribution $n_i(p)$ of the harmonic-oscillator (HO) shell-model single-particle state $i$, the effects of \emph{NN} correlations that give widths to the energy distributions of the whole strengths in ($e,e'$) and ($e,e'p$) reactions were studied in detail in~\cite{PhysRevC.83.045504}. The conclusion was that, with a symmetric energy spread for the single-particle levels, it is not possible to get an asymmetry of the longitudinal scaling function similar to that observed by the data. The next step was to use single-particle momentum distributions that correspond to natural orbitals (NO's) for the single-particle wave functions and occupation numbers using a method where short-range NN correlations are taken into account. As known, the NO's $\varphi_\alpha (r)$ are defined~\cite{PhysRev.97.1474} as the complete orthonormal set of single-particle wave functions that diagonalize the one-body density matrix (OBDM) $\rho(\mathbf{r},\mathbf{r'})$:
\begin{equation}
\rho (\mathbf{r},\mathbf{r}^{\prime} )=\sum_{a} N_{a} \varphi_{a}^{*}(\mathbf{r}) \varphi_{a}
(\mathbf{r}^{\prime}) ,
\label{defNO}
\end{equation}
where the eigenvalues $N_{\alpha}$ ($0\leq N_{\alpha}\leq 1$, $ \sum_{\alpha} N_{\alpha}=A$) are the natural occupation numbers. In~\cite{PhysRevC.83.045504} the OBDM obtained within the lowest-order approximation of the Jastrow correlation method~\cite{PhysRevC.48.74} has been used. Though the use of NO's enhances the value of the maximum and slightly reduces the tails of the scaling functions, it leads to a weak asymmetry of the scaling function $f(\psi)$ that is not in accordance with the significant tail extended to positive $\psi$-values, seen in the analysis of the data of the ($e,e'$) process. At the same time, the usage of NO's leads to significant high-momentum tail of the momentum distribution in contrast to the case in MFA approaches. An important conclusion reached in~\cite{PhysRevC.83.045504} was that the strong asymmetry of the scaling function $f(\psi)$ at positive $\psi$ values (observed by the analysis of data) emerges when FSI (and other peculiarities of the electron scattering beyond the PWIA) are taken into account.

\subsection{FSI\label{sec:2.2}}

The analyses of the FSI in the case of inclusive electron-nuclei scattering performed in Ref.~\cite{PhysRevC.83.045504} (following Ref.~\cite{PhysRevC.77.044311}) concerned two types of FSI effects, the Pauli blocking and the interaction of the struck nucleon with the spectator system described by means of the time-independent optical potential (OP):
\begin{equation}\label{OP}
    U=V-\imath W\, .
\end{equation}
The latter (see Ref.~\cite{PhysRevC.22.1680}) can be accounted for by replacing in the PWIA expression for the  inclusive electron-nucleus cross section:
\begin{multline}\label{cr.s.}
        \frac{d\sigma_t}{d\omega d\n q}={2\pi\alpha^2}\frac{\n q}{E_{\ve k}^2}
        \int dE\:d^3p\:\frac{S_t(\ve p, E)}{E_{\ve p}E_{\ve {p'}}}\times\\
\times        \delta\big(\omega+M-E-E_\ve{p'}\big)L_{\mu\nu}^\text{em}H^{\mu\nu}_{\text{em, }t}\,
\end{multline}
the energy-conserving delta-function by
\begin{equation}\label{deltaf}
 \delta (\omega+M-E-E_\ve{p'}) \rightarrow \dfrac{W/\pi}{W^2+[\omega+M-E-E_\ve{p'}-V]^2}.
\end{equation}
In Eq.(\ref{cr.s.}) the index $t$ denotes the nucleon isospin, $L_{\mu\nu}^{\text{em}}$ and $H^{\mu\nu}_{\text{em, }t}$ are the leptonic and hadronic tensor, respectively, and $S_t(\ve p, E)$ is the proton (neutron) spectral function. The quantities $E_{\ve k}$, $E_{\ve p}$, $E_{\ve {p'}}$ and $E$ are the initial electron energy, the energy of the nucleon inside the nucleus, the energy of the ejected nucleon, and the removal energy (see Ref.~\cite{PhysRevC.22.1680} for  details). The real ($V$) and imaginary ($W$) parts of the OP in Eqs.~(\ref{OP}) and (\ref{deltaf}) are obtained in Ref.~\cite{Cooper:1993nx} from the Dirac OP. Spatially averaged values of these OP components, evaluating them at the $r$ values that match their respective root mean-squared radii~\cite{Cooper:1993nx} have been used in  Ref.~\cite{PhysRevC.83.045504}. Finally, the OP $U(p')$ related to the scalar ($S$) and vector ($V$) parts of the potential in~\cite{Cooper:1993nx} is obtained in the form (see also Ref.~\cite{PhysRevC.77.044311}):
\begin{equation}\label{opf}
    E_\ve{p'}+U(\ve{p'})=\sqrt{[M+S(T_\ve{p'},\bar r_S)]^2+\ve{p'}^2}+V(T_\ve{p'},\bar r_V).
\end{equation}

Alternatively, in Ref.~\cite{PhysRevC.83.045504} also an OP with the following imaginary part of the potential $U(p')$ (given in Ref.~\cite{Nakamura2005201}) was considered:
\begin{equation}\label{opW}
W=\dfrac{\hbar c}{2}\rho_\text{nucl}\sigma_{NN}\dfrac{|\mathbf{p}'|}{E_{\mathbf{p}'}} \, ,
\end{equation}
with particular values of $\rho_\text{nucl}$ and $\sigma_{NN}$ for $^{16}$O nucleus. In the present work we restrict ourselves to the first approach [Eq.~(\ref{opf})].

\subsection{Scaling functions and charge-changing neutrino-nucleus reaction cross section\label{sec:2.3}}

In this subsection we follow the description of the formalism concerning the charge-changing (CC) (anti)neutrino-nucleus cross section given in Ref.~\cite{PhysRevC.71.015501} (see also Ref.~\cite{PhysRevC.74.054603}). The CC neutrino cross section in the target laboratory frame is given in the form:
\begin{equation}
\left [ \frac{d^{2}\sigma}{d\Omega dk^{\prime}}\right
]_{\chi}\equiv \sigma_{0}{\cal F}_{\chi}^{2},
\label{c.c.cr.s.}
\end{equation}
where $\chi=+$ for neutrino-induced reaction (\emph{e.g.}, $\nu_{\ell}+n\rightarrow \ell^{-}+p$, where $\ell=e, \mu, \tau$) and $\chi=-$ for antineutrino-induced reactions (\emph{e.g.}, $\overline{\nu}_{\ell}+p\rightarrow \ell^{+}+n$),
\begin{equation}
\sigma_{0}\equiv
\frac{(G\cos\theta_{c})^{2}}{2\pi^{2}}[k^{\prime}\cos\tilde{\theta}/2]^{2},
\label{eq:80}
\end{equation}
$G=1.16639\times 10^{-5}$ GeV$^{-2}$ being the Fermi constant, $\theta_{c}$ being the Cabibbo angle  $(\cos\theta_{c}=0.9741)$, and
\begin{equation}
\tan^{2}\tilde{\theta}/2\equiv \frac{|Q|^{2}}{v_{0}},
\label{eq:81}
\end{equation}
\begin{equation}
v_{0}\equiv (\epsilon +
\epsilon^{\prime})^{2}-q^{2}=4\epsilon\epsilon^{\prime}-|Q|^{2}.
\label{eq:82}
\end{equation}

In Eqs.~(\ref{c.c.cr.s.})-(\ref{eq:82}) $\Omega$, $k'$ and $\epsilon'$ are the scattering angle, momentum and energy of the outgoing lepton. The quantitiy ${\cal F}_{\chi}^{2}$ in Eq.~(\ref{c.c.cr.s.}) depends on the nuclear structure and it is presented in Ref.~\cite{PhysRevC.71.015501} as a generalized Rosenbluth decomposition having charge-charge,
charge-longitudinal, longitudinal-longitudinal, and two types of transverse responses ($R$'s). These nuclear response functions are expressed in terms of the nuclear tensor $H^{\mu\nu}$ in the QE (as well as in the $\Delta$-region) by means of its relationships with the scaling functions from the particular model used. In the calculations of the $\nu$-nucleus cross sections the Galster parametrization~\cite{Galster1971221} of the form factors in the vector sector was used, whereas in the axial-vector sector the form factors given in  Ref.~\cite{PhysRevC.71.015501} were used.

In the present work we evaluate the double-differential cross section for CCQE (anti)neutrino induced process averaged over the neutrino flux $\Phi(\epsilon_\nu)$:
\begin{equation}
\frac{d^2\sigma}{dT_\mu d\cos\theta_\mu}
= \frac{1}{\Phi_\text{tot}} \int
\left[ \frac{d^2\sigma}{dT_\mu d\cos\theta_\mu} \right]_{\epsilon_\nu}
\Phi(\epsilon_\nu) d\epsilon_\nu ,
\end{equation}
where $T_\mu$ and $\theta_\mu$ are the kinetic energy and the scattering angle of the outgoing muon, respectively, $\epsilon_\nu$ is the neutrino energy and $\Phi_\text{tot}$ is the total integrated $\nu_\mu$ flux factor for the MiniBooNE experiment. The results of the calculations will be presented as a function of the muon kinetic energy $T_\mu$ and as a function of the scattering angle $\theta_\mu$. The results obtained by integrating the flux-averaged double-differential cross sections over the angle:
\begin{equation}
\left\langle\frac{d\sigma}{dT_\mu }\right\rangle
= \frac{1}{\Phi_\text{tot}} \int\Phi(\epsilon_\nu)\int
\left[ \frac{d^2\sigma}{dT_\mu d\cos\theta_\mu} \right]_{\epsilon_\nu}d\cos\theta_\mu
 d\epsilon_\nu ,
\end{equation}
as well as those obtained by integrating over the muon kinetic energy
\begin{equation}
\left\langle\frac{d\sigma}{d\cos\theta_\mu }\right\rangle
= \frac{1}{\Phi_\text{tot}} \int\Phi(\epsilon_\nu)\int
\left[ \frac{d^2\sigma}{dT_\mu d\cos\theta_\mu} \right]_{\epsilon_\nu}dT_\mu
 d\epsilon_\nu ,
\end{equation}
will be presented as well.

Finally, the results of the calculations of the total cross sections of CCQE (anti)neutrino scattering from $^{12}$C will be given as a function of the (anti)neutrino energy and compared with the existing experimental data. The calculations of the cross sections mentioned above are performed using different models to evaluate the spectral function, namely RFG, NO and HO, and accounting also for the role of the FSI, correspondingly RFG+FSI, NO+FSI and HO+FSI.

\section{Results and discussion \label{sec:3}}

\begin{figure}[t]
\includegraphics[width=0.95\columnwidth]{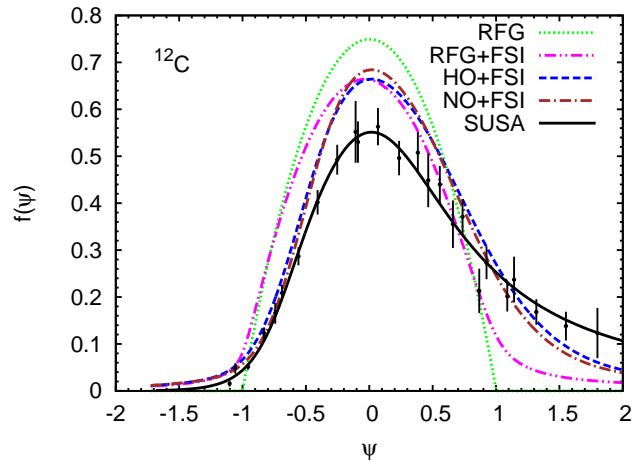}
\caption{(Color online) Results for the scaling function $f (\psi)$ for $^{12}$C obtained using RFG+FSI, HO+FSI, and NO+FSI approaches are compared with the RFG and SUSA results, as well as with the longitudinal experimental data~\cite{Jourdan1996117}.}\label{fig00}
\end{figure}

\begin{figure*}[t]
\includegraphics[width=.66\columnwidth]{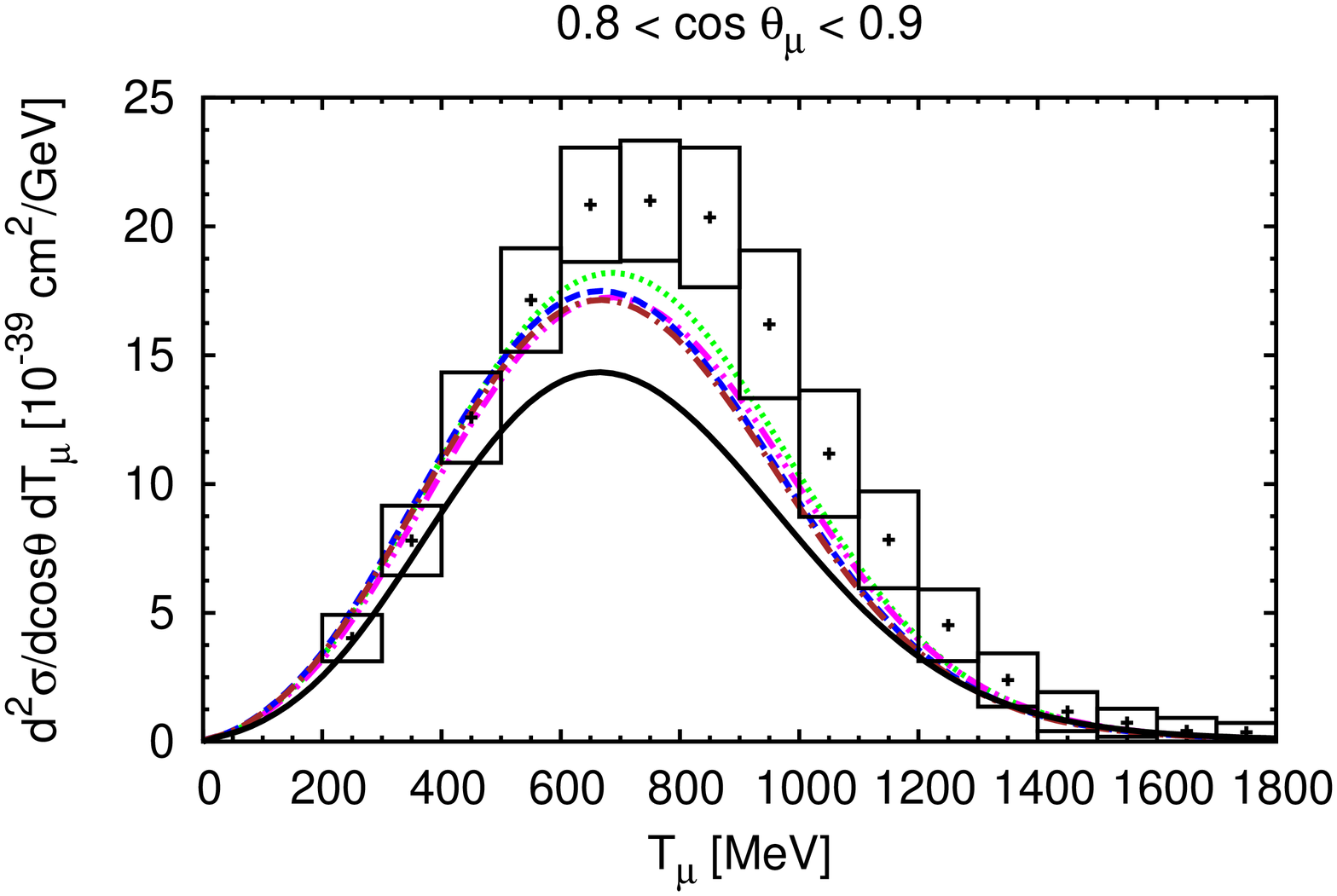}\includegraphics[width=.66\columnwidth]{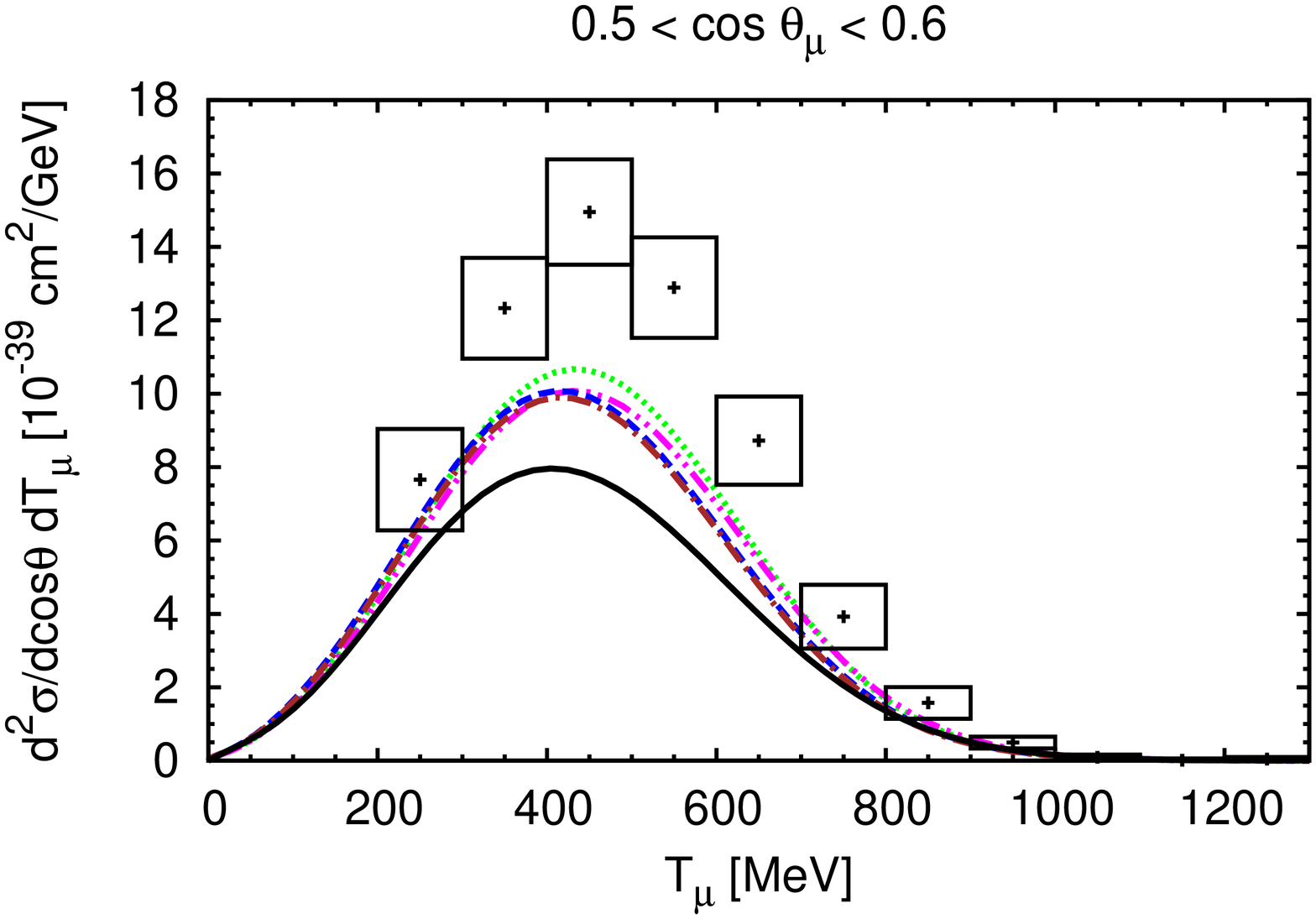}\includegraphics[width=.66\columnwidth]{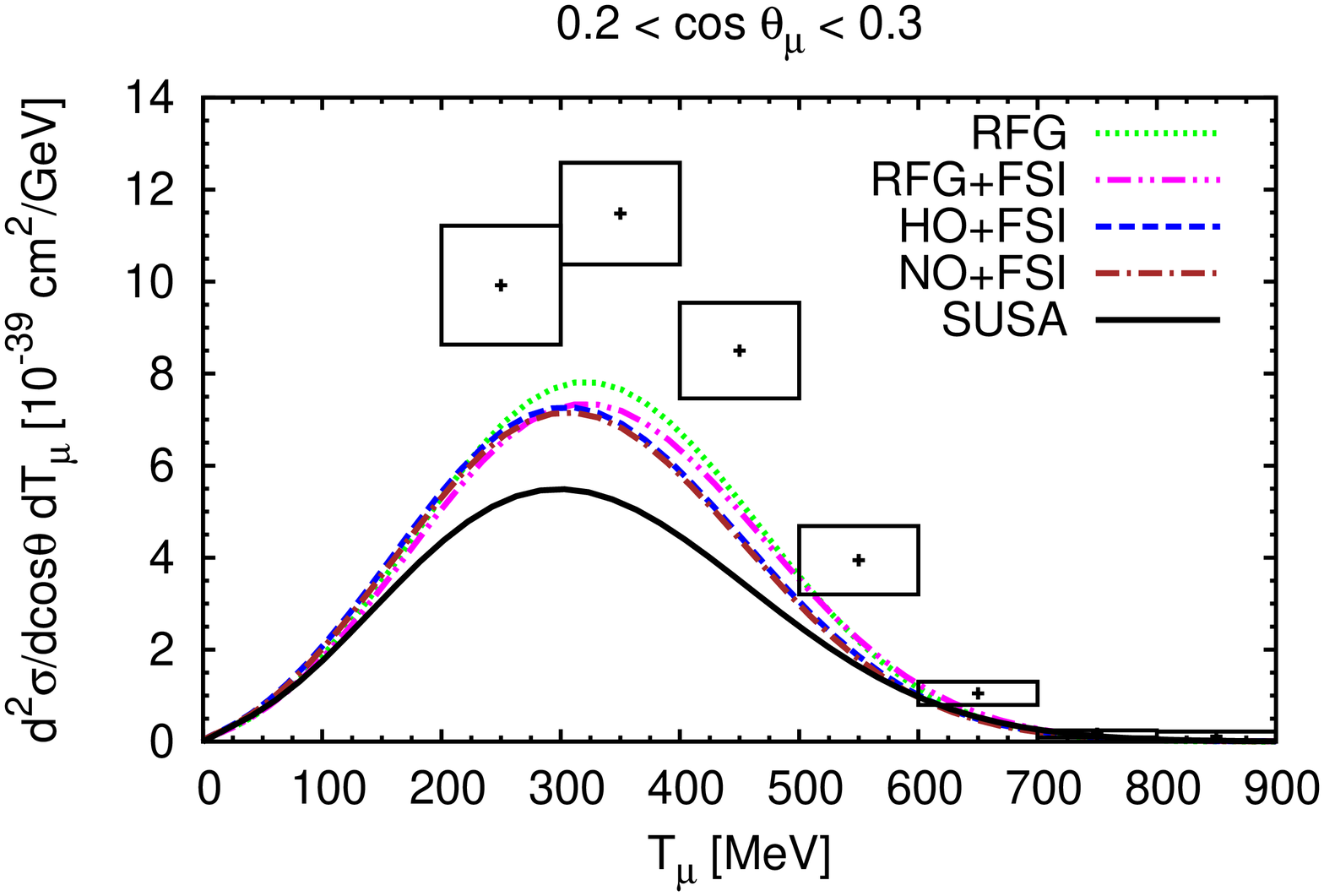}\\[5pt]
\includegraphics[width=.66\columnwidth]{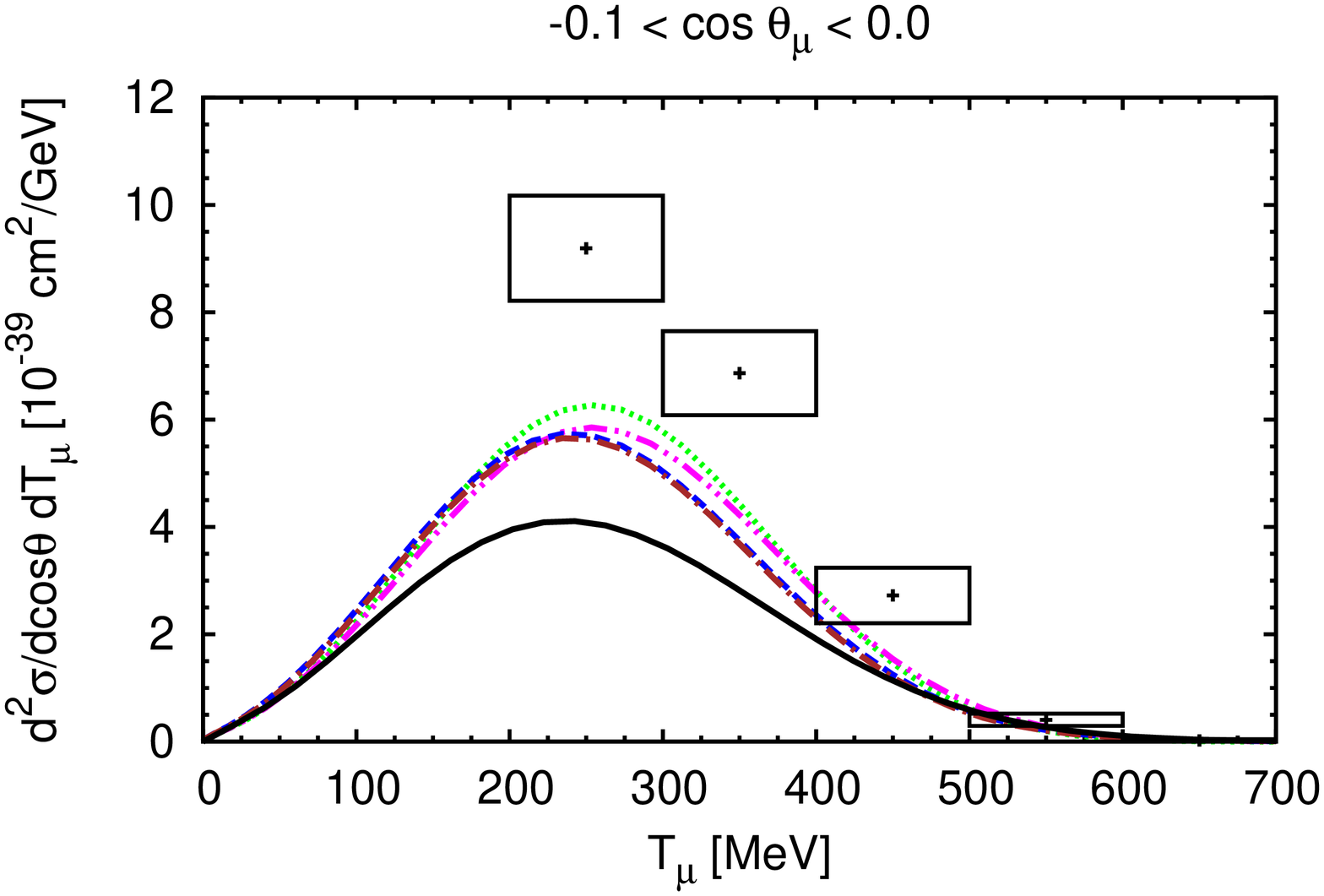}\includegraphics[width=.66\columnwidth]{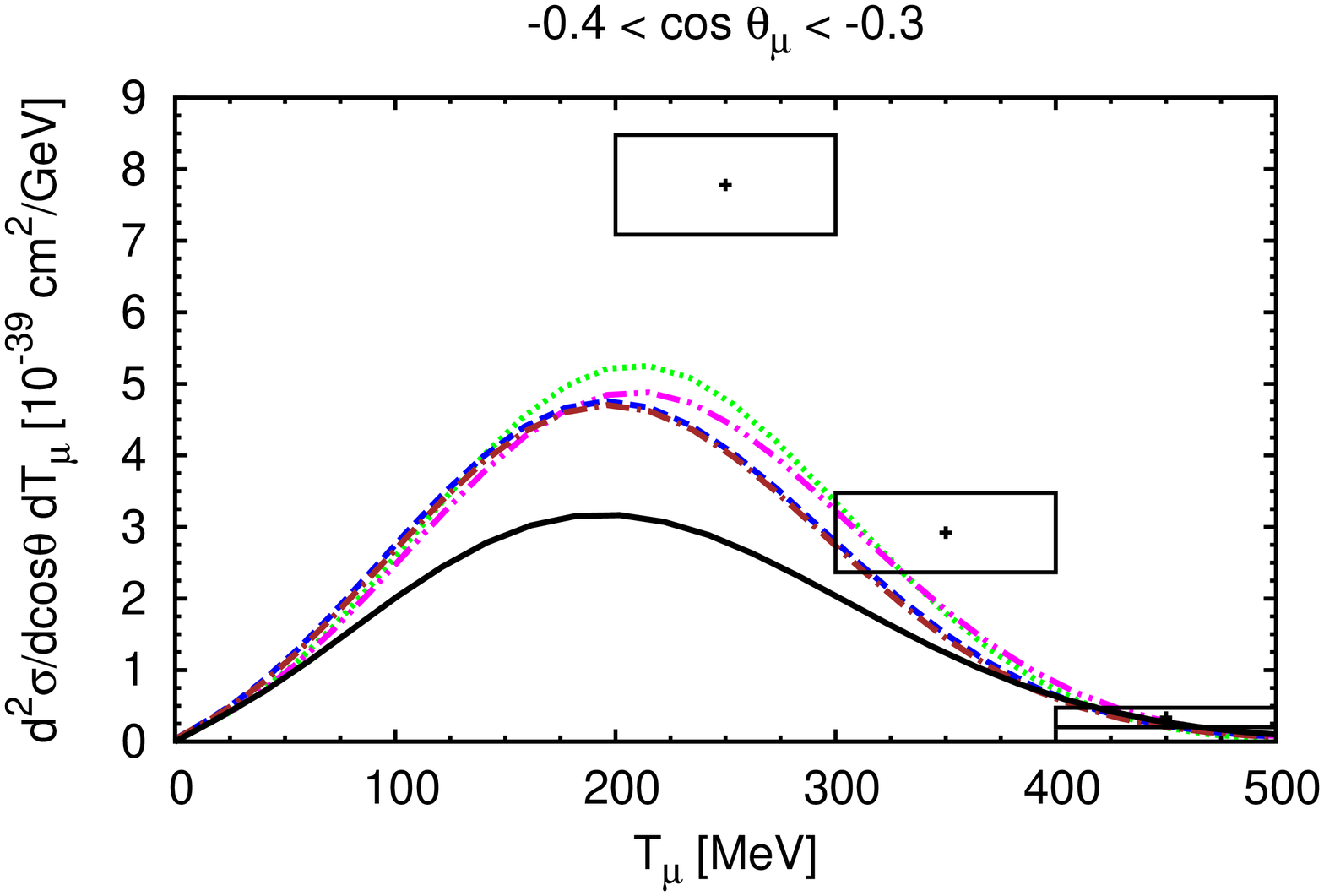}\includegraphics[width=.66\columnwidth]{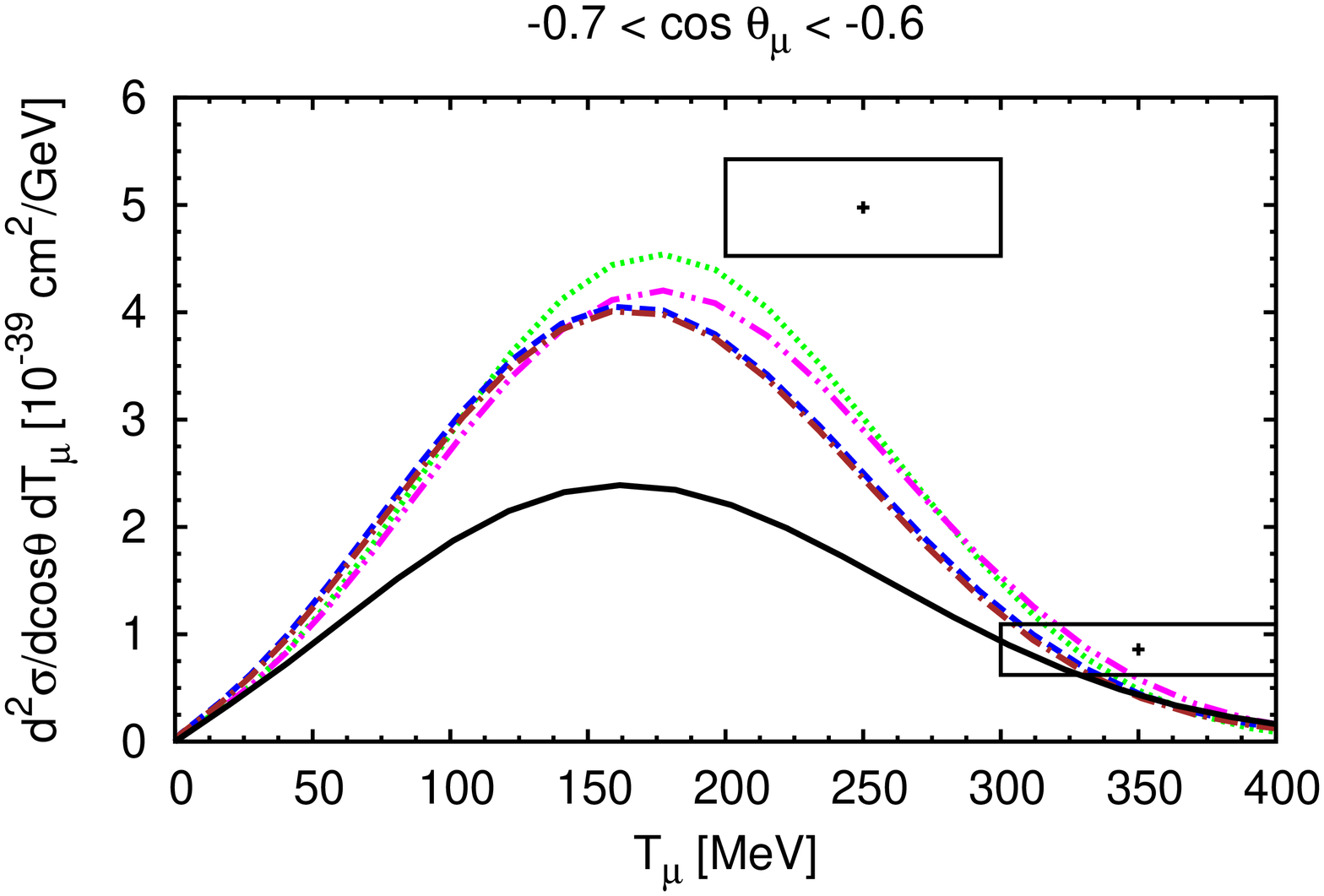}
\caption{(Color online) Flux-integrated double-differential cross section per target nucleon for the $\nu_\mu$ CCQE process on $^{12}$C displayed versus the $\mu^-$ kinetic energy $T_\mu$ for various bins of $\cos\theta_\mu$ obtained within the RFG+FSI, HO+FSI, and NO+FSI approaches for $M_A=1.03$~GeV. The data are from Ref.~\cite{miniboone}.}\label{fig01}
\end{figure*}

\begin{figure*}[t]
\includegraphics[width=.66\columnwidth]{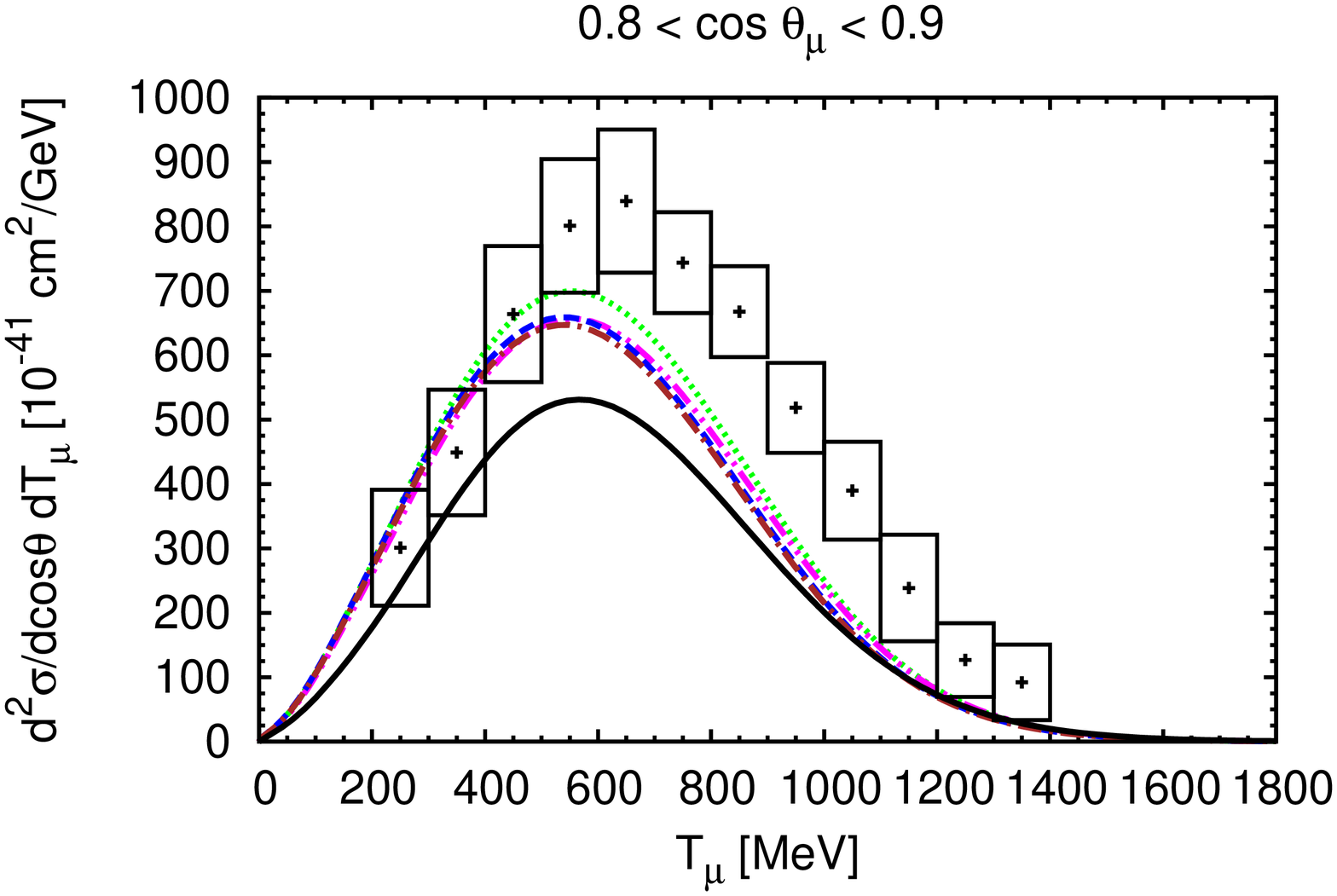}\includegraphics[width=.66\columnwidth]{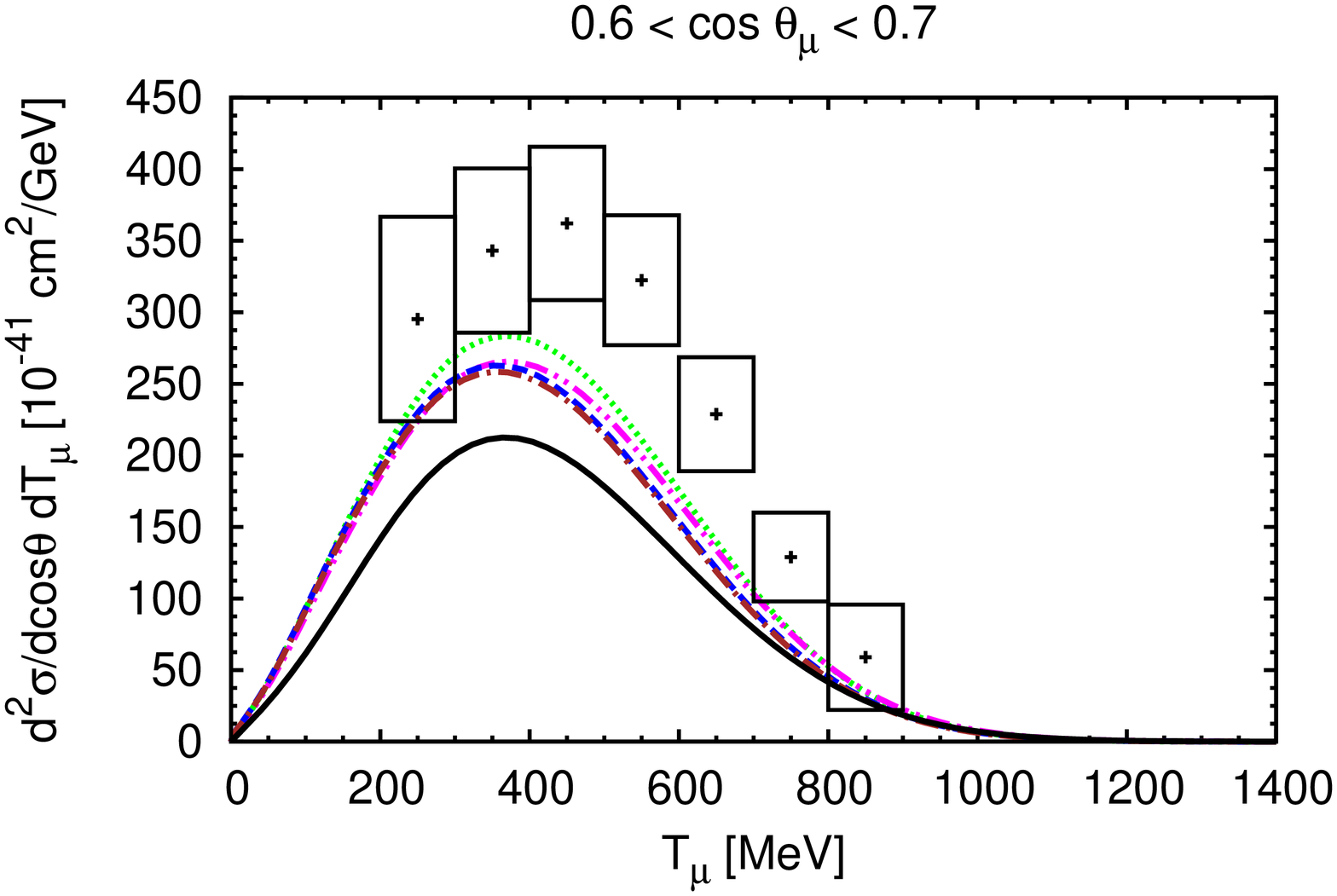}\includegraphics[width=.66\columnwidth]{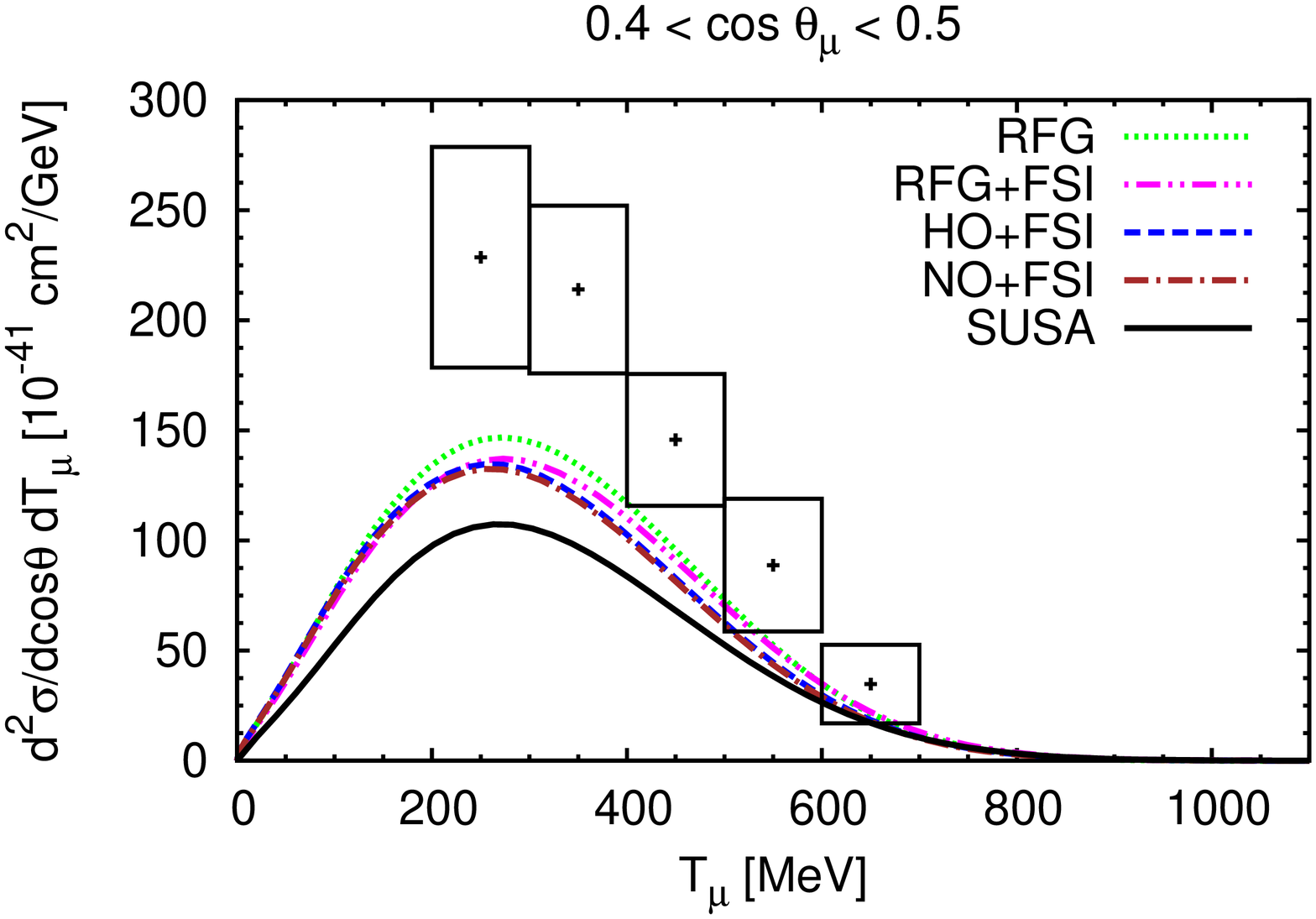}\\[5pt]
\includegraphics[width=.66\columnwidth]{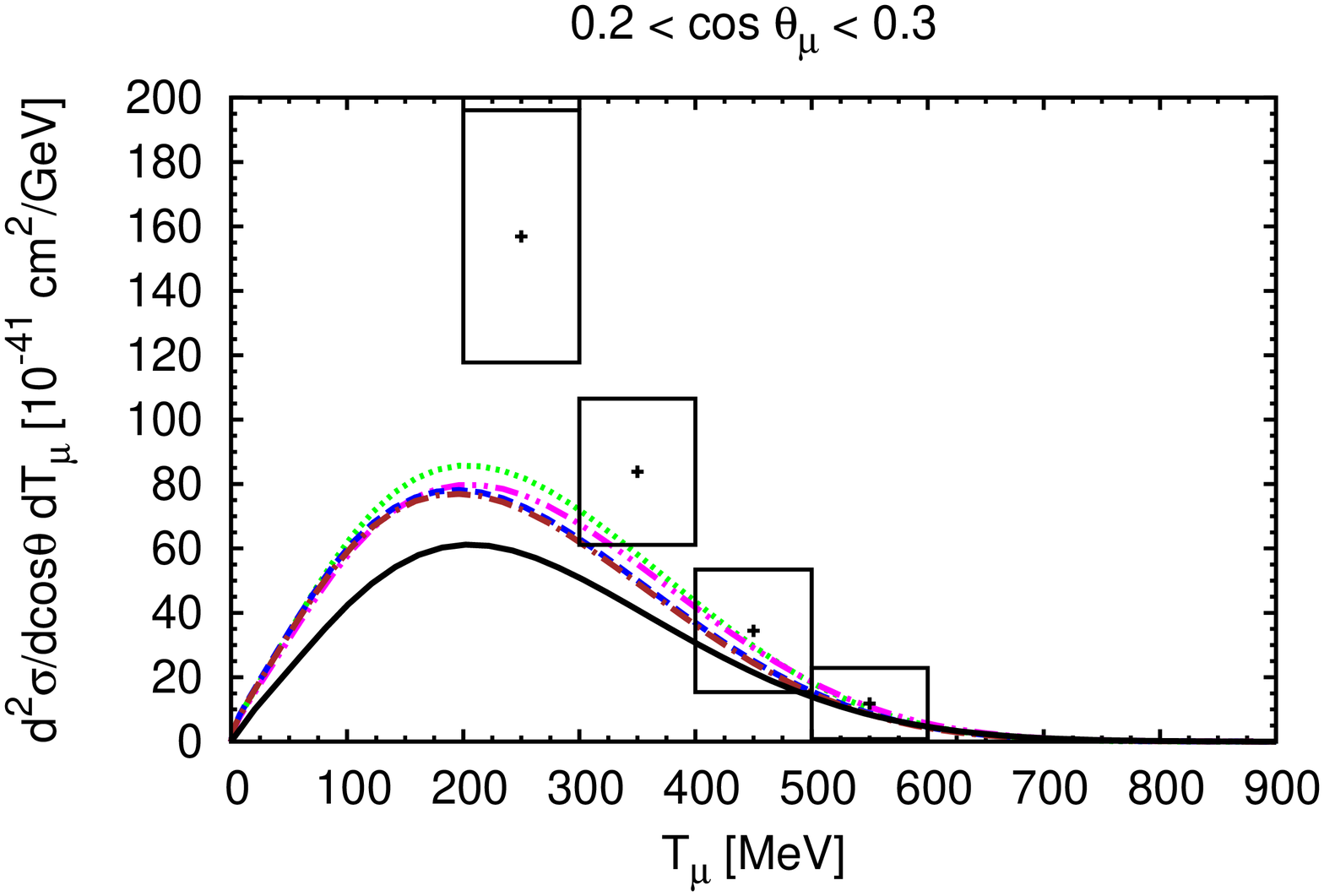}\includegraphics[width=.66\columnwidth]{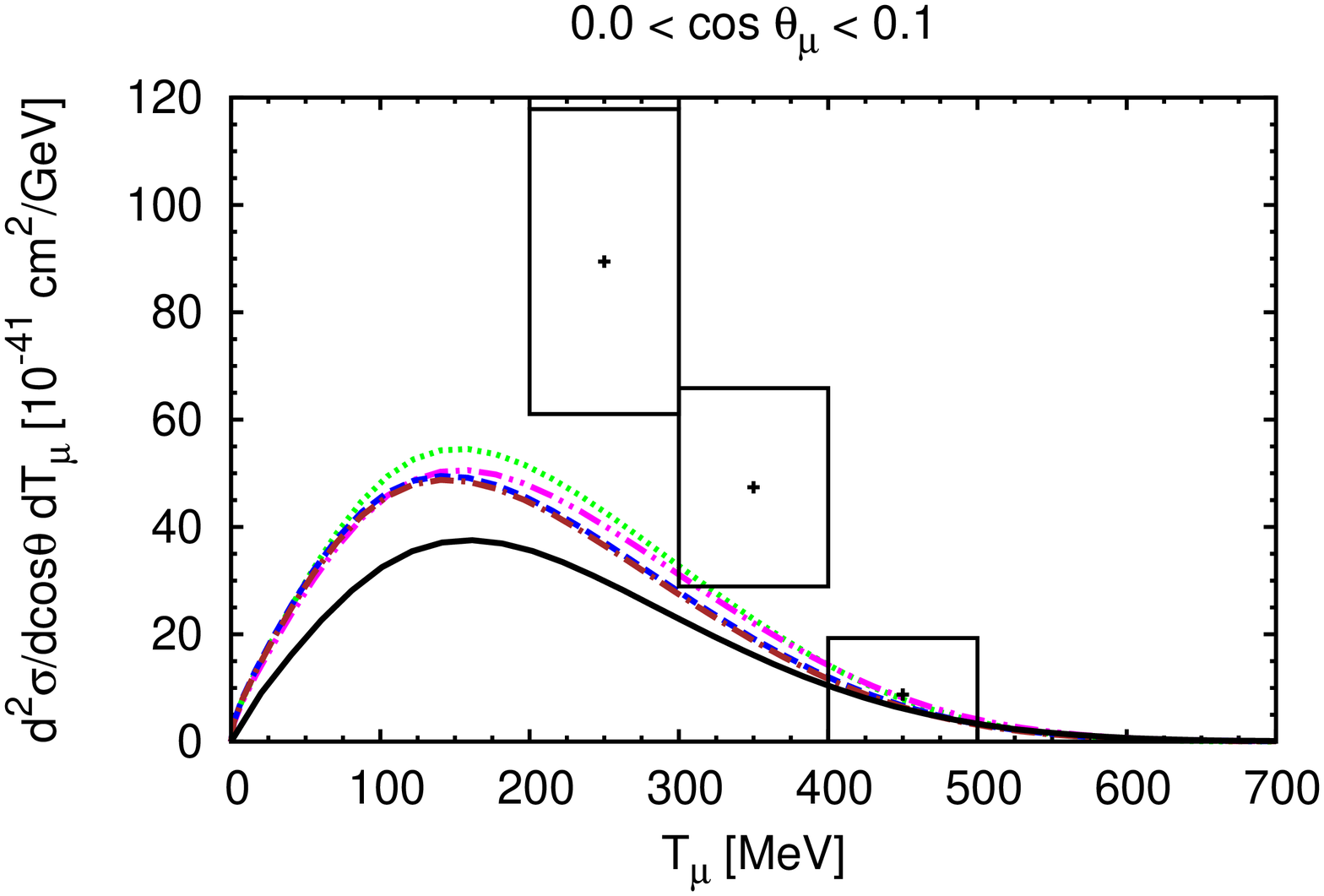}\includegraphics[width=.66\columnwidth]{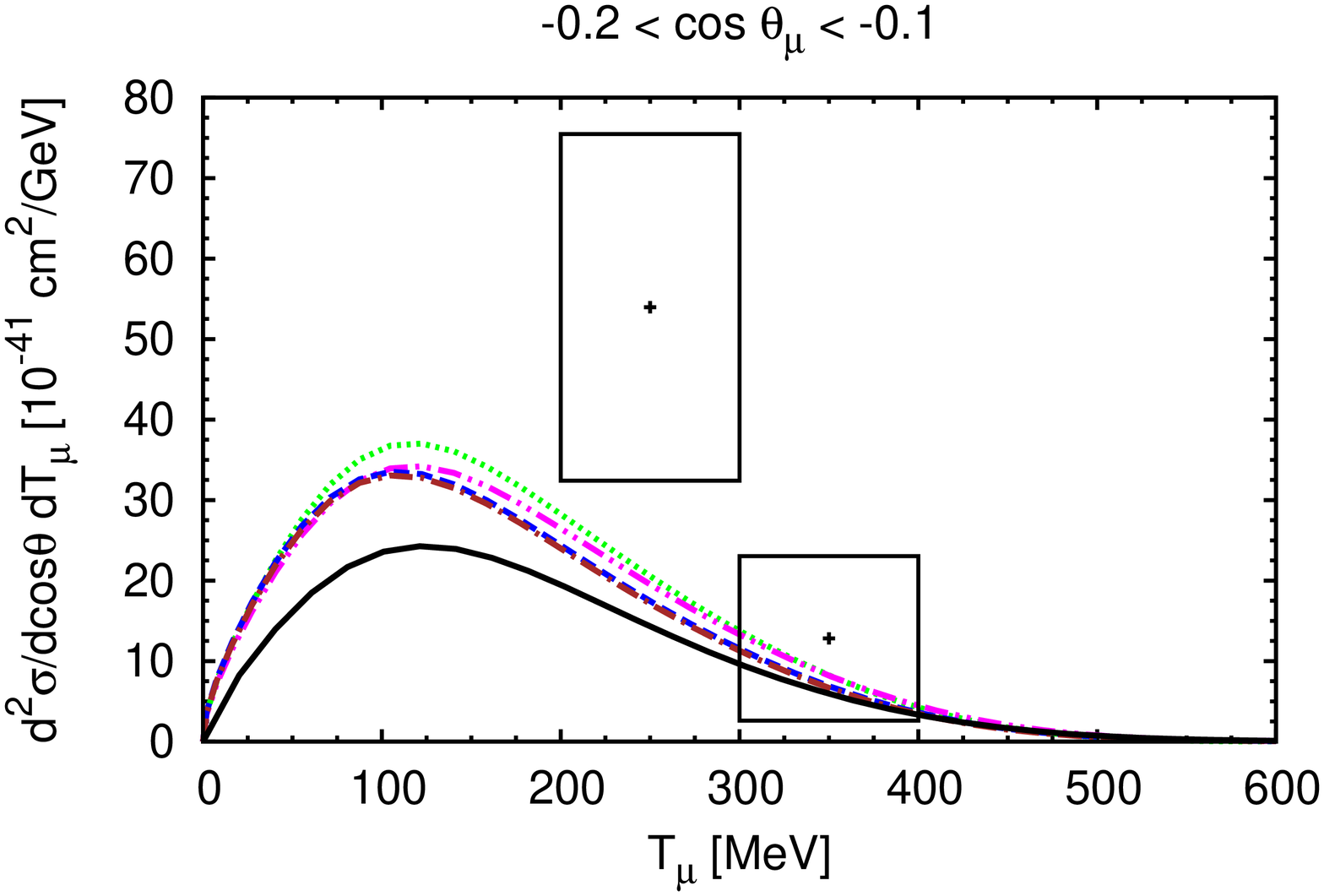}
\caption{(Color online) As for Fig.~\ref{fig01}, but for $\overline\nu_\mu$ scattering versus $\mu^+$ kinetic energy $T_\mu$. The data are from Ref.~\cite{miniboone-ant}.}\label{fig02}
\end{figure*}

\begin{figure*}[t]
\includegraphics[height=40mm]{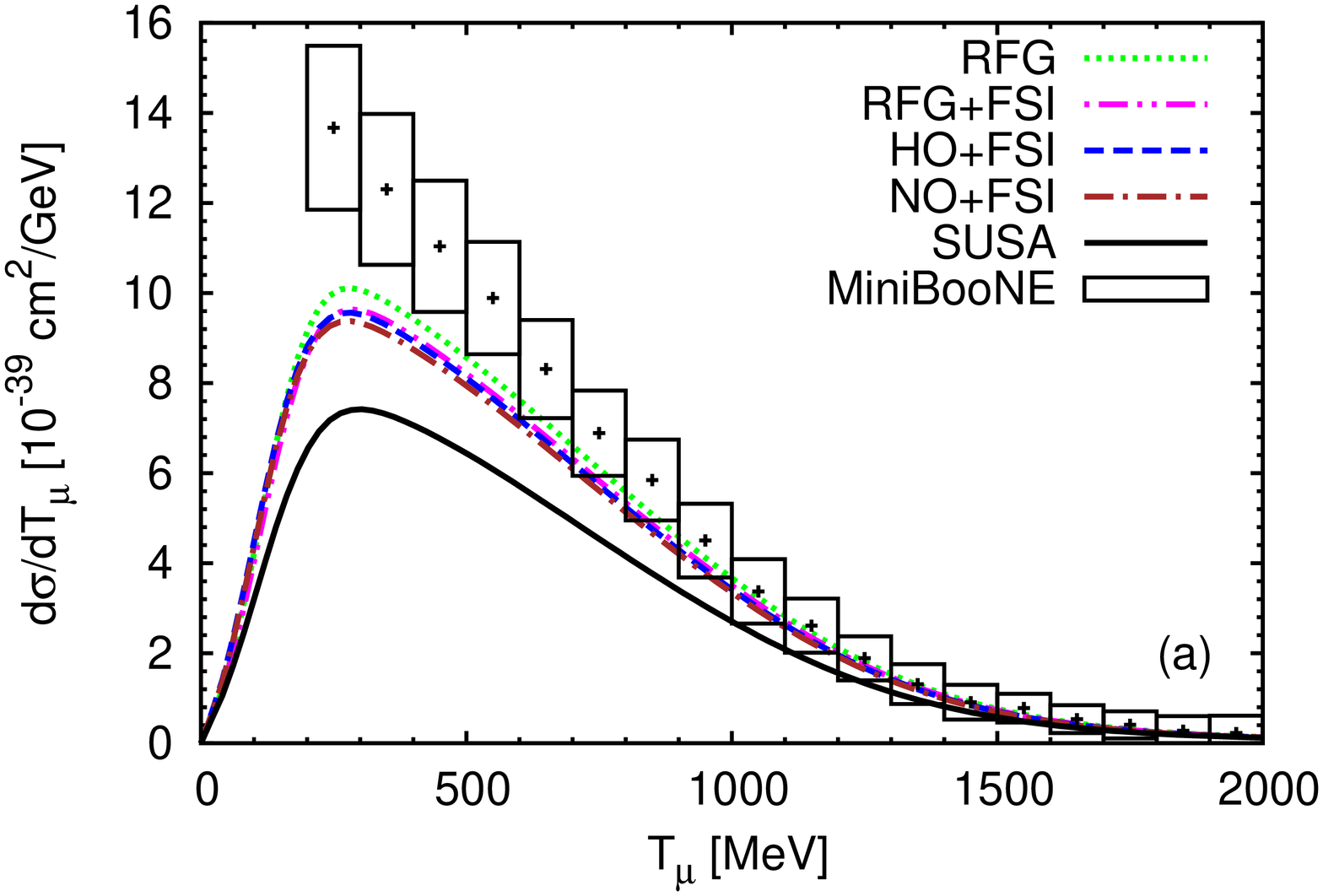}\includegraphics[height=40mm]{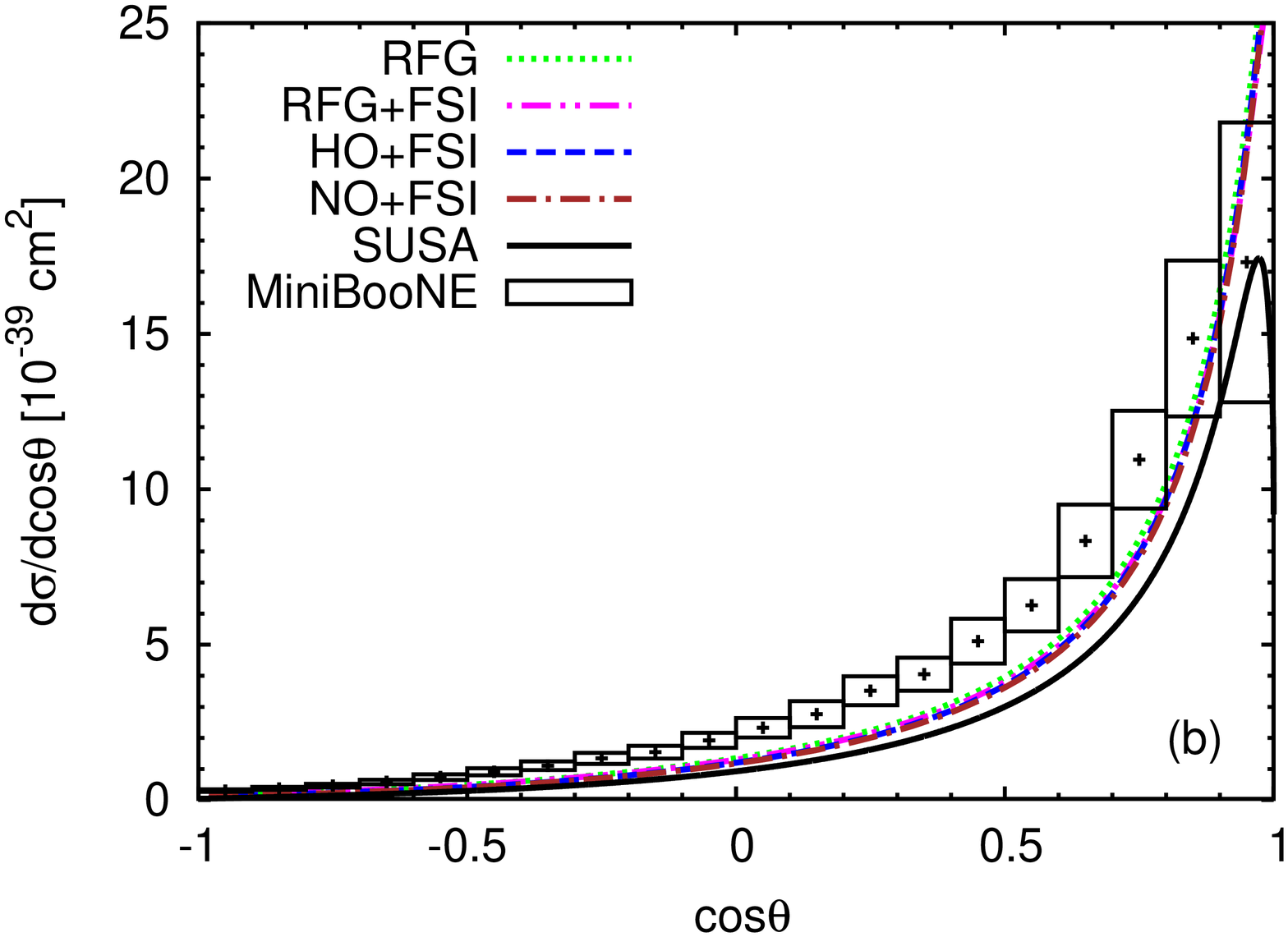}\includegraphics[height=40mm]{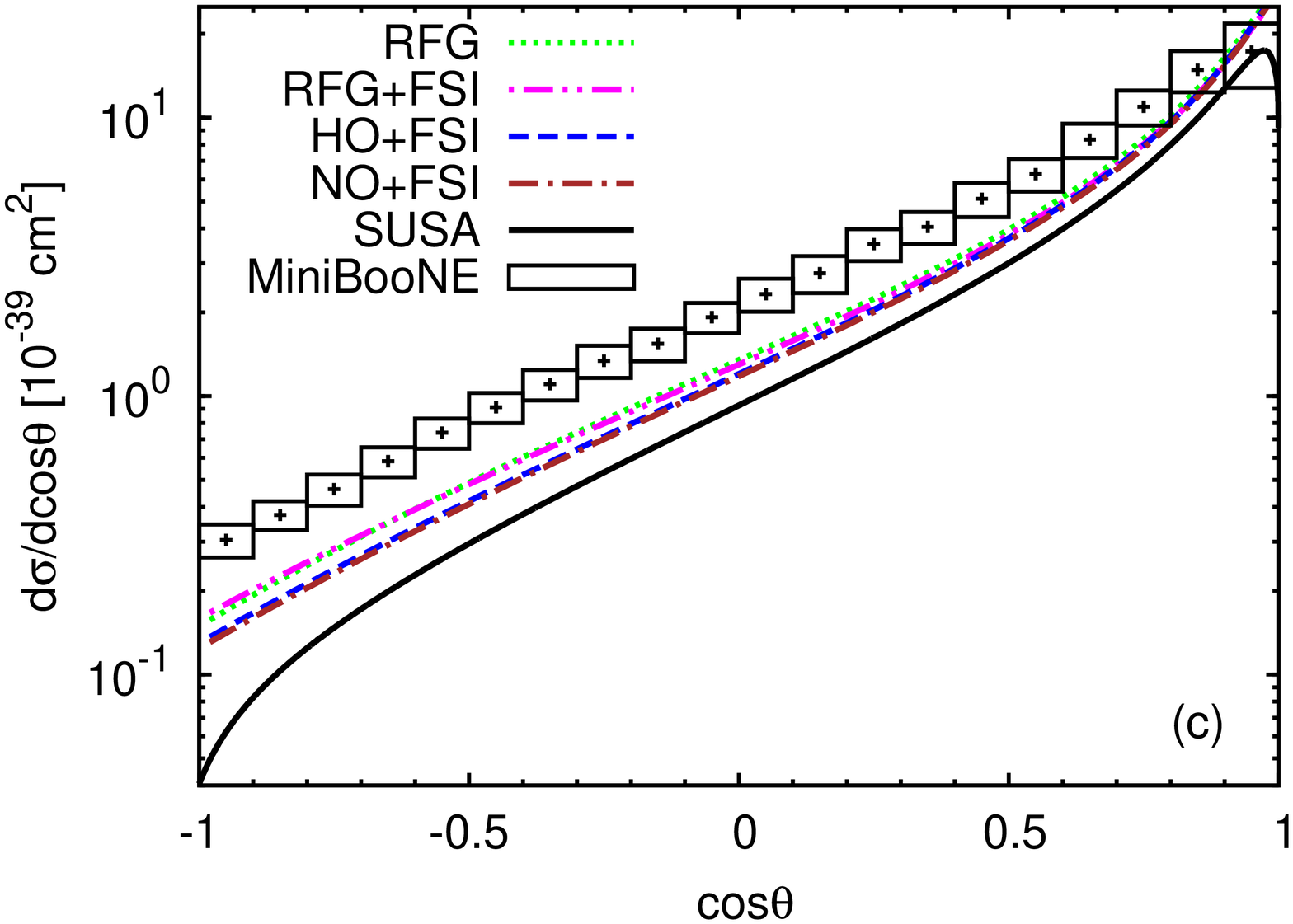}\\[5pt]
\includegraphics[height=40mm]{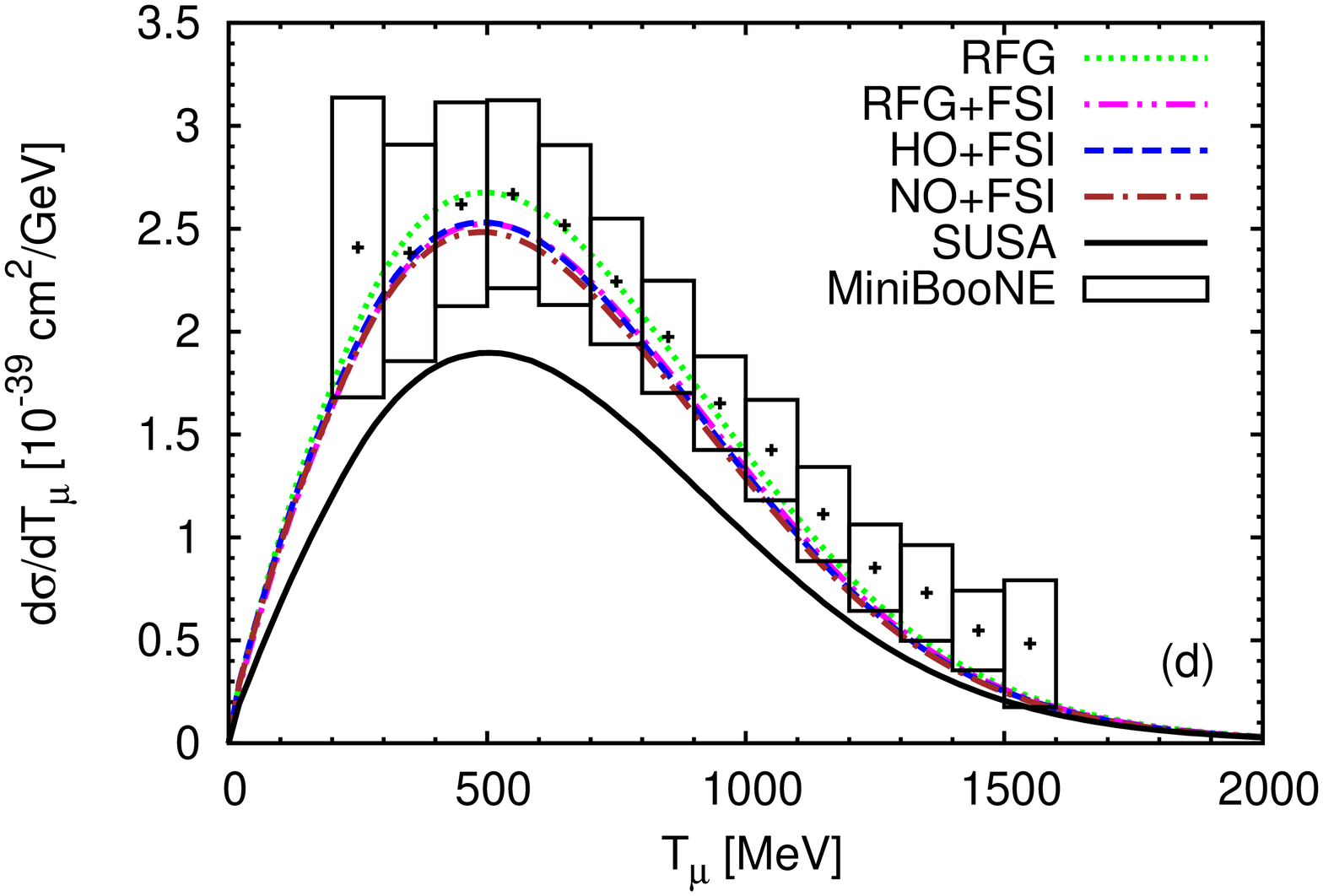}\includegraphics[height=40mm]{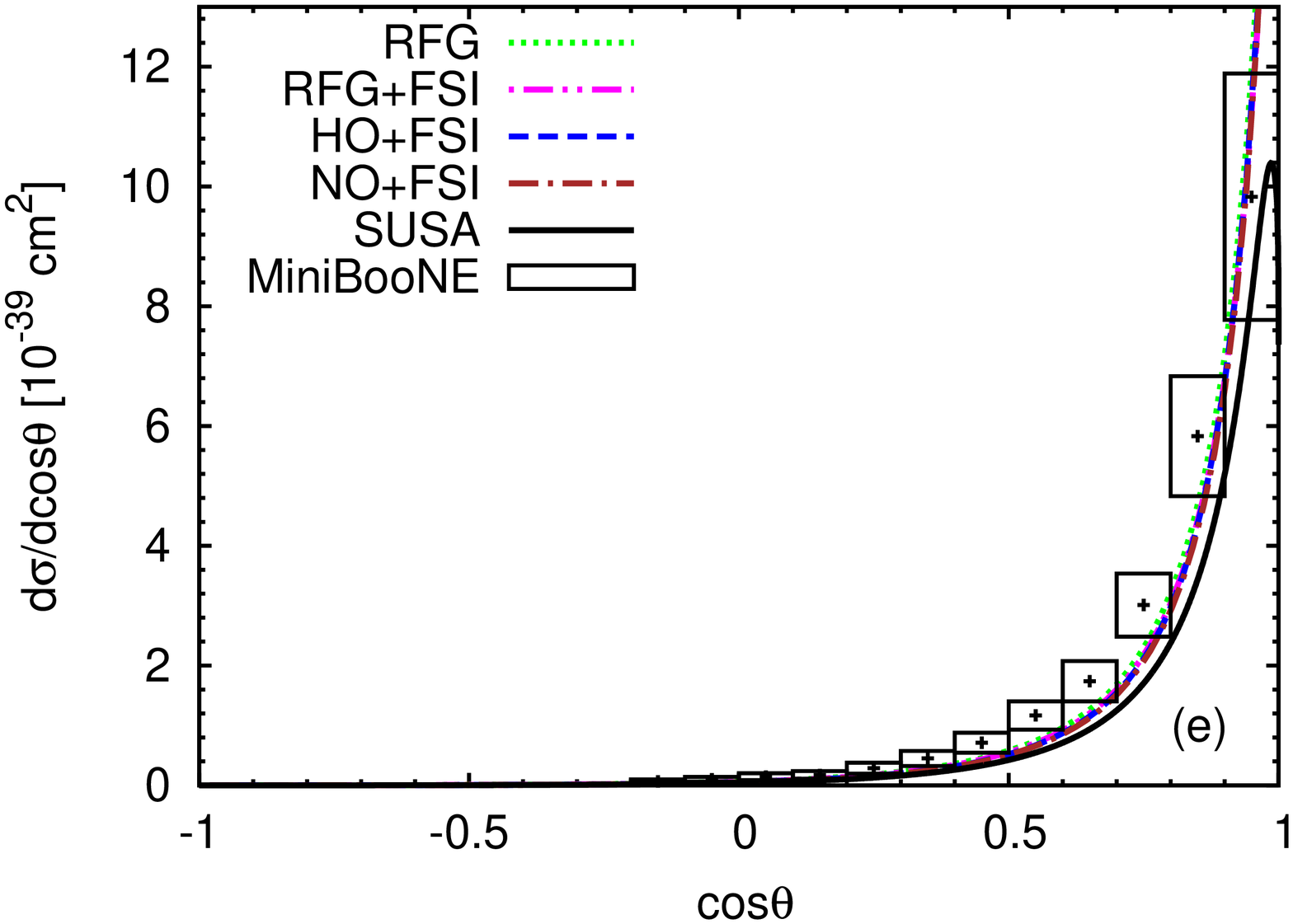}\includegraphics[height=40mm]{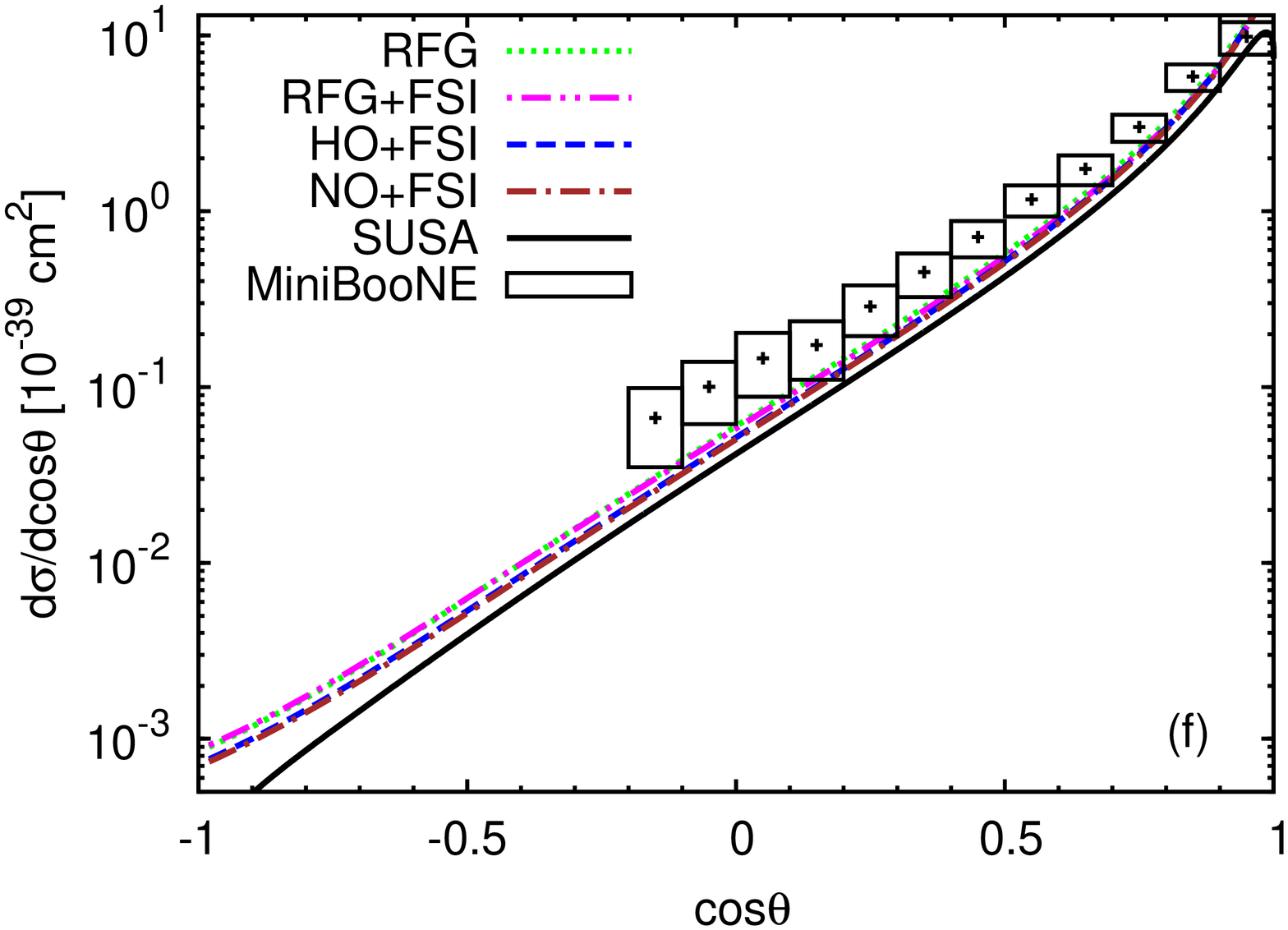}
\caption{(Color online) MiniBooNE flux-averaged CCQE $\nu_\mu$-$^{12}$C differential cross section per nucleon as a function: (a) of the muon kinetic energy, (b) of the muon scattering angle (normal scale), (c) of the muon scattering angle (logarithmic scale). MiniBooNE flux-averaged CCQE $\overline{\nu}_\mu$-$^{12}$C differential cross section per nucleon are given in bottom panels [(d)--(f)]. The data are from Ref.~\cite{miniboone,miniboone-ant}.}\label{fig03}
\end{figure*}

\begin{figure}[t]
\includegraphics[width=.85\columnwidth]{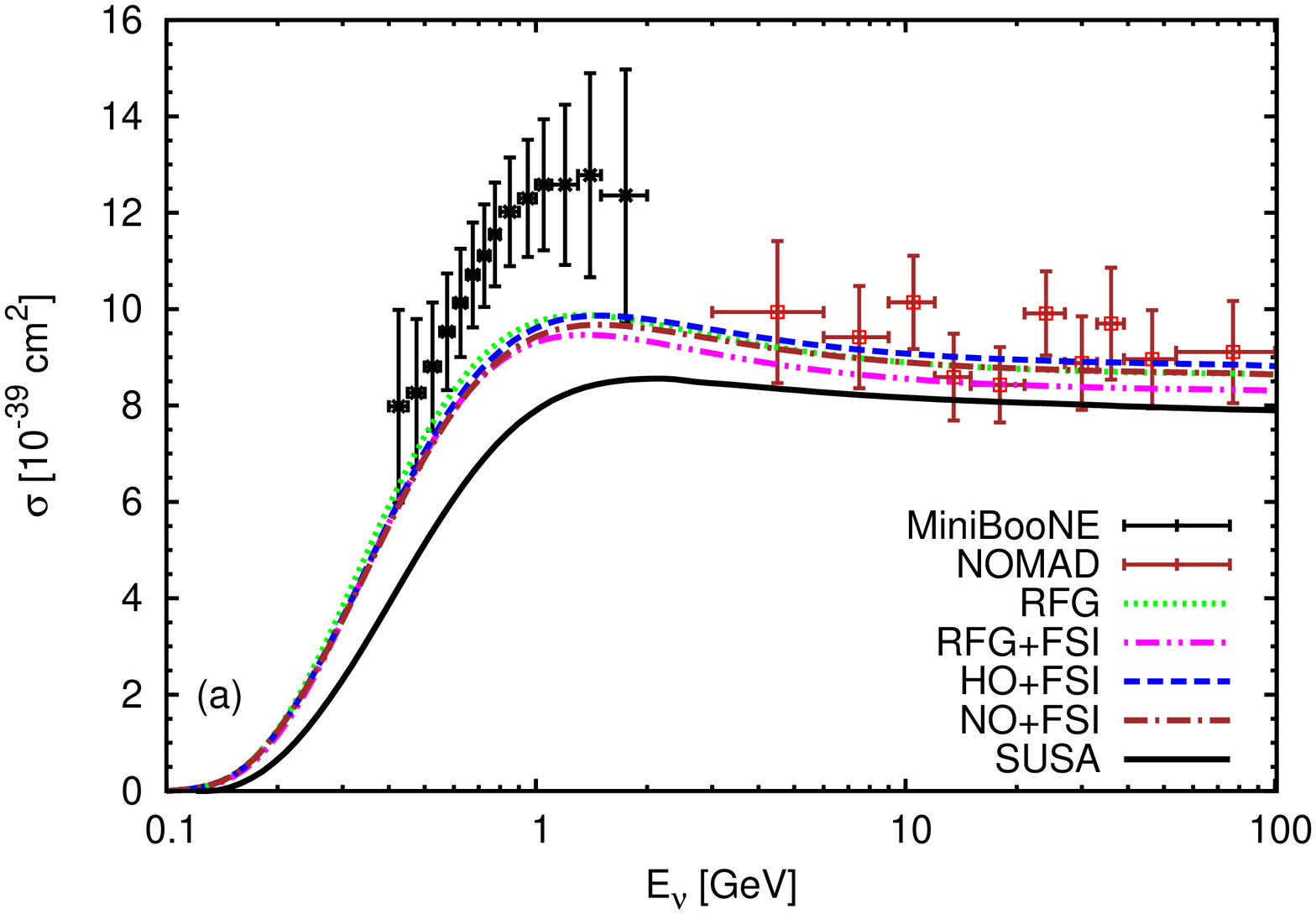}\\[5pt]
\includegraphics[width=.85\columnwidth]{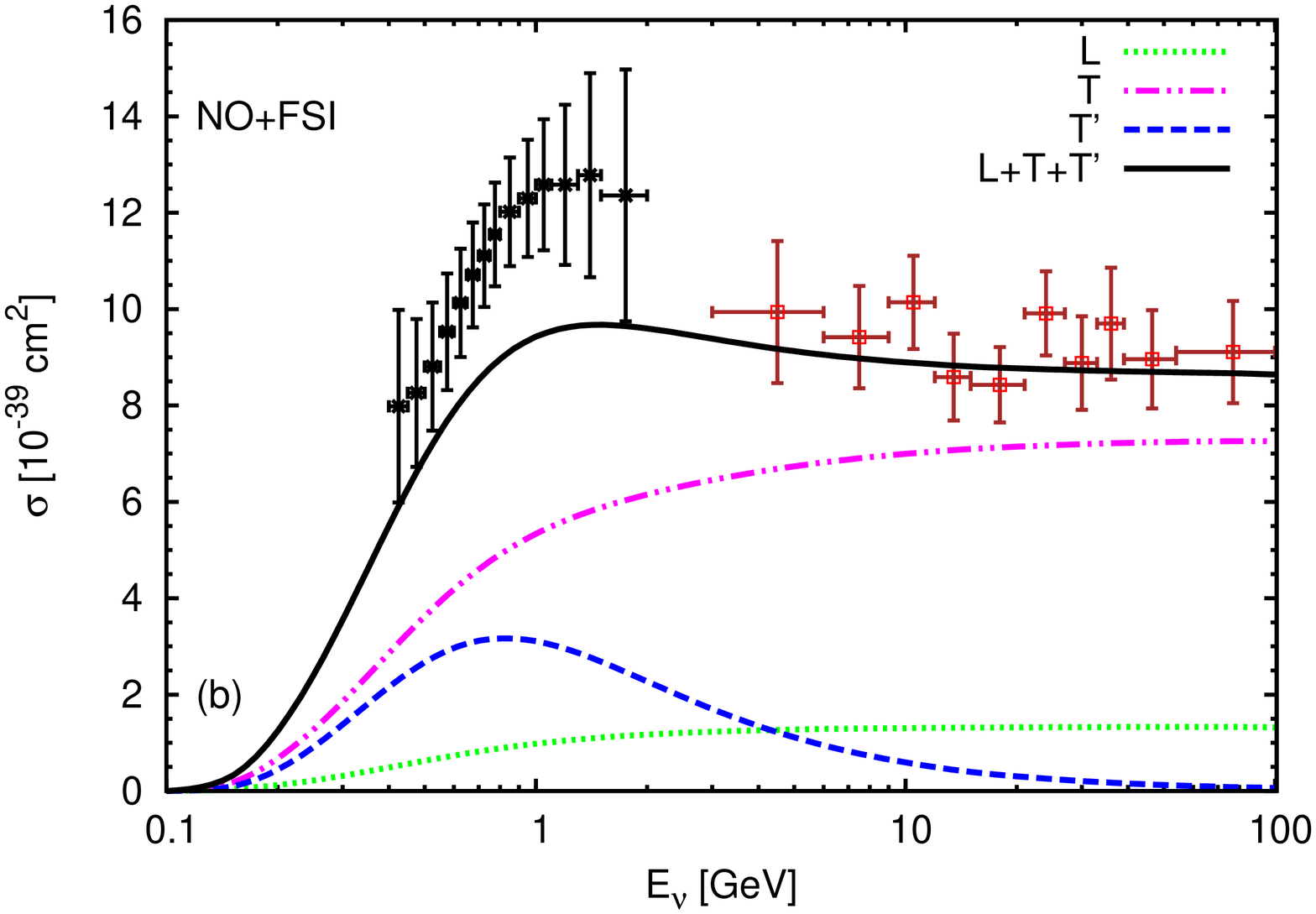}\\[5pt]
\includegraphics[width=.85\columnwidth]{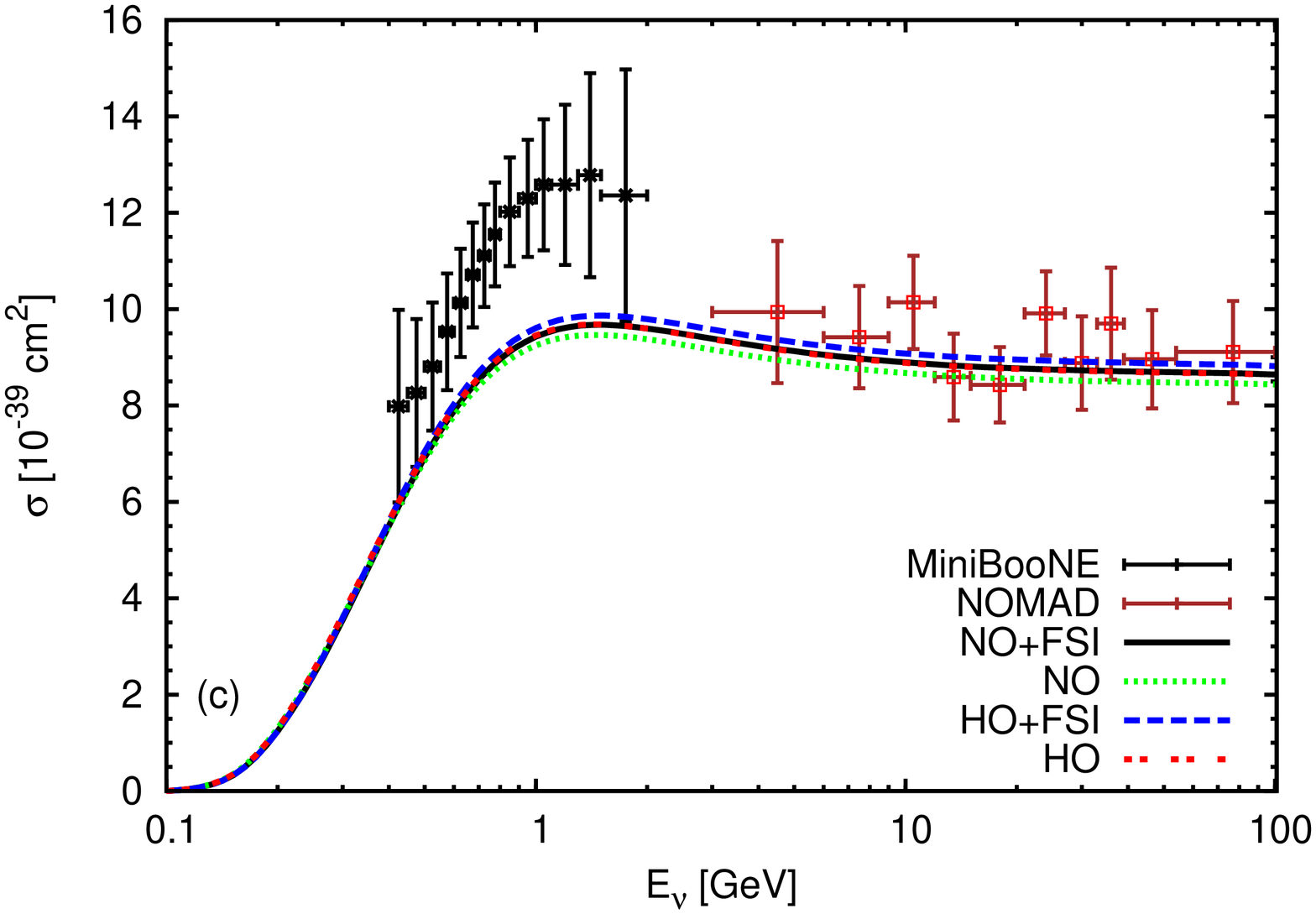}
\caption{(Color online) (a) CCQE $\nu_\mu-^{12}$C cross sections per nucleon displayed versus neutrino energy $E_\nu$ and evaluated using the RFG+FSI, HO+FSI, and NO+FSI approaches with the standard value of the axial-vector dipole mass $M_A = 1.03$~GeV/c$^2$ are compared with the MiniBooNE~\cite{miniboone} and NOMAD~\cite{lyubushkin_study_2009} experimental data. (b) Separated contributions in the NO+FSI approach. (c) The cross sections within HO and NO approaches with and without accounting for FSI.}\label{fig04}
\end{figure}

\begin{figure}[t]
\includegraphics[width=.9\columnwidth]{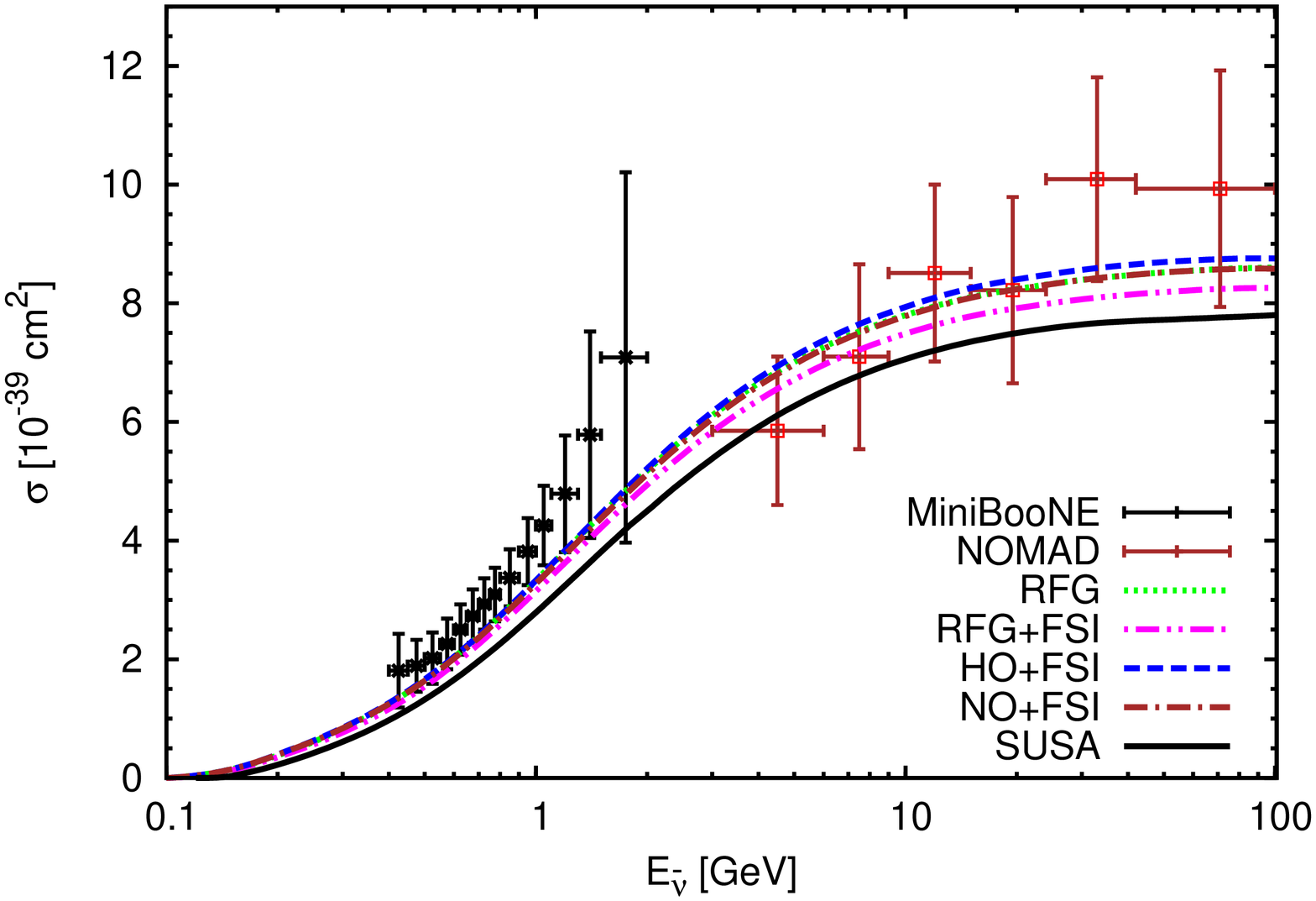}
\caption{(Color online) As in Fig.~\ref{fig04}, but for $\overline{\nu}_\mu-^{12}$C scattering. The MiniBooNE data are from~\cite{miniboone-ant}.}\label{fig05}
\end{figure}

In this section we present firstly (in Fig.~\ref{fig00}) our results for the scaling function $f(\psi)$ using the relationships given in subsection~\ref{sec:2.1} and taking into account the FSI as described in  subsection~\ref{sec:2.2}. The procedure for the calculations of the scaling function $f(\psi)$ is the following:
\begin{itemize}
\item[(i)] The spectral function $S(p,{\cal E})$ is constructed in the form of Eq.~(\ref{HF+lorent});
\item [(ii)] The single-particle momentum distributions $n_i(p)$ are taken to be either corresponding to the HO single-particle wave functions or to the NO’s from the Jastrow correlation method;
\item [(iii)] The Lorentzian function [Eq.~(\ref{lorent})] is used for the energy dependence of the spectral function with parameters $\Gamma_{1p} = 6$~MeV, $\Gamma_{1s} = 20$~MeV, which are fixed to the experimental widths of the $1p$ and $1s$ states in $^{12}$C ~\cite{Dutta1999};
\item[(iv)] For a given momentum transfer $q = 1$~GeV/c and energy of the initial electron $\varepsilon = 1$~GeV we calculate the electron-nucleus ($^{12}$C) cross section by using Eq.~(\ref{cr.s.}) in which the spectral function $S(p,{\cal E})$ [Eq.~(\ref{HF+lorent})] is used;
\item[(v)] The corresponding scaling function $F(q,\omega)$ is calculated within the PWIA by means of Eq.~(\ref{scaling}) and by multiplying it by $k_A$ the scaling function $f(\psi)$ is obtained;
\item[(vi)] To account for FSI, the $\delta$-function in Eq.~(\ref{cr.s.}) is replaced by Eq.~(\ref{deltaf})) and Eq.~(\ref{opf}) is used.
\end{itemize}
In this way the results for the HO+FSI (dashed line) and NO+FSI (dash-dotted line) are obtained. As a reference are shown also the scaling functions in the cases of SuSA (solid line), RFG (dotted line) and RFG+FSI (two dots-dashed line). In the RFG case the corresponding RFG spectral function is used in Eq.~(\ref{cr.s.}), and in the RFG+FSI case the energy conserving $\delta$-function in Eq.~(\ref{cr.s.}) is replaced by Eq.~(\ref{deltaf}). As can be seen from Fig.~\ref{fig00}, accounting for FSI leads to a small asymmetry of the scaling function, see \emph{e.g.}, the comparison between the RFG and the RFG+FSI scaling functions. We found that the asymmetry in the scaling function gets larger by using the Lorentzian function [Eq.~(\ref{lorent})] for the energy dependence of the spectral function (in the HO+FSI and NO+FSI cases) than by using the Gaussian function [Eq.~(\ref{eq3a})].

The flux-integrated double-differential cross section per target nucleon for the $\nu_\mu$ ($\overline{\nu}_\mu$) CCQE process on $^{12}$C displayed versus the $\mu^-$ ($\mu^+$) kinetic energy $T_\mu$ for various bins of $\cos\theta_\mu$ (both forward and backward angles) obtained within the RFG+FSI, NO+FSI and HO+FSI approaches are given in Fig.~\ref{fig01} (Fig.~\ref{fig02}). In all calculations the standard value of the axial mass $M_A= 1.032$~GeV/c$^2$ is used. We emphasize that, as expected, our results underpredict the data due to the fact that our calculations are based on the impulse approximation (IA). However, the shape of the cross section is reproduced by the RFG+FSI, HO+FSI, and NO+FSI approaches. The results lie close to each other due to the small differences between the scaling functions. We note that as the angle increases the curves corresponding to the HO+FSI and NO+FSI cases (whose scaling functions are more asymmetric) deviate from the RFG+FSI ones. The same behavior can be seen in Fig.~\ref{fig03}, where we present results of MiniBooNE flux-averaged CCQE  $\nu_\mu$($\overline{\nu}_\mu$)-$^{12}$C single-differential cross section per nucleon as a function of the muon kinetic energy [(a) and (d)] and of the muon scattering angle [(b), (c), (e), and (f)]. In Figs.~\ref{fig03}(c) and \ref{fig03}(f) are displayed differential cross sections as functions of the muon scattering angle in logarithmic scale in order to make the difference between the cross sections for the negative and positive values of $\cos\theta_\mu$ more visible.

In Fig.~\ref{fig04}(a) the total cross sections obtained within RFG+FSI, NO+FSI and HO+FSI approaches are presented. The calculations are performed up to $100$~GeV for comparison with the NOMAD experimental data~\cite{lyubushkin_study_2009}. All models give results that agree with the NOMAD data but underpredict the MiniBooNE ones, more seriously in the neutrino than in the antineutrino case. Also, in Fig.~\ref{fig04}(b) the results for the pure vector-transverse ($T$), longitudinal ($L$), and axial-transverse ($T'$) contributions to the cross sections within NO+FSI approach are  presented. In the next Fig.~\ref{fig05} we present our antineutrino results that are in a good agreement with the NOMAD data. We note that the comparison with the MiniBooNE data in the case of antineutrino scattering (Fig.~\ref{fig05}) shows much better agreement than in the neutrino case (Fig.~\ref{fig04}). As can be seen from Fig.~\ref{fig04}(b), the maximum of the axial-transverse ($T'$) contribution is around the maximum of the neutrino flux at the MiniBooNE experiment. The effects of $T'$ contributions to the cross sections are negligible at energies above $10$~GeV.

As observed, for very high $\nu_\mu$ ($\overline{\nu}_\mu$) energies (above $\sim 10$~GeV) the total cross section for neutrinos and antineutrinos is very similar. This is consistent with the negligible contribution given by the $T'$ response in this region. Only the $L$ and $T$ channels contribute for the higher values explored by NOMAD experiment (where the theory is in accordance with data). On the contrary, in the region explored by the MiniBooNE collaboration, the main contributions come from the two transverse $T$, $T'$ channels, being constructive (destructive) in neutrino (antineutrino) cross sections. As already mentioned, effects beyond the IA, {\it i.e.,} $2p-2h$ MEC, may have a significant contribution in the transverse responses leading to theoretical results closer to data. However, note that the enhancement needed to fit data should be larger for neutrinos than for antineutrinos, hence a careful analysis of $2p-2h$ MEC contributions in both transverse responses is needed before more definite conclusions can be drawn.

The results we obtained using realistic spectral functions without FSI are in qualitative good agreement with those of Ref.~\cite{PhysRevLett.105.132301} (with standard axial mass used) and Ref.~\cite{Ankowski:2012ei}, where a realistic hole spectral function for $^{12}$C, obtained in the local density approximation, was used. The inclusion of FSI does not dramatically change the results, but gives a slight redistribution of the strength in the differential cross sections and a small depletion of the integrated cross section in the case of RFG approach. Using more realistic SF (HO or NO) effects of FSI leads to small increase of the integrated cross sections, as can be seen in Fig.~\ref{fig04}(c) due to the larger tails of scaling functions at negative and positive values of $\psi$.

Concluding this Section we emphasize that no calculations based on the spectral function is able to reproduce the MiniBooNE data. The discrepancy is most likely due to missing of the effects beyond the impulse approximation, \emph{e.g.} those of the $2p-2h$ MEC that have contribution in the \emph{transverse responses}. This concerns also the similar disagreement with theory that appears when the phenomenological scaling function in SuSA is used. The latter is a \emph{purely longitudinal quasielastic response} extracted from inclusive electron scattering data and thus is nearly insensitive to $2p-2h$ MEC contributions.

\section{Conclusions \label{sec:4}}

In the present work we give results for the CCQE (anti)neutrino cross sections on a $^{12}$C target and compare them with the available data from the MiniBooNE~\cite{miniboone,miniboone-ant} and the  NOMAD~\cite{lyubushkin_study_2009}
Collaborations. The results presented and discussed are for: flux-integrated double-differential cross section per target nucleon versus the muon kinetic energy and versus the muon scattering angle; the  $\nu_\mu$($\overline{\nu}_\mu$)$+ ^{12}$C total cross sections as a function of the (anti)neutrino energy. The method we use is based on a spectral function $S(p,{\cal E})$ that gives a scaling function in  accordance with the ($e,e'$) scattering data. The spectral function: i) accounts for short-range \emph{NN} correlations by using natural orbitals from the Jastrow correlation method to obtain the single-particle momentum distributions $n_i(p)$ that are ingredients of $S(p,{\cal E})$, ii) it has a realistic energy dependence using parameters that are fixed to the experimental widths of $1p$ and $1s$ states in $^{12}$C, iii) it is used also in extending the range of the (anti)neutrino energy from the analysis of the MiniBooNE experimental data to the NOMAD data as a step to clarify the limits of the superscaling approaches. The results are compared with those when \emph{NN} correlations are not included, \emph{e.g.}, in the RFG model and when HO single-particle wave functions are used instead of NO's in the calculations of $n_i(p)$. The effects of FSI are accounted for following the approaches from Refs.~\cite{PhysRevC.77.044311,PhysRevC.22.1680,Cooper:1993nx} and the results (NO+FSI, HO+FSI, RFG+FSI) are compared with those without FSI. Also, a comparison with SuSA results is presented. In all calculations we use the standard value of the axial mass $M_A= 1.032$~GeV/c$^2$.

Our main findings can be summarized as follows:
\begin{itemize}
\item[1.] The use of different spectral functions (RFG, HO, NO) gives quite similar (within $5-7\%$) CCQE neutrino cross sections at all energies, signalling that the process is not too sensitive to the specific treatment of the bound state.
\item[2.] The effect of FSI is a depletion of the cross section of about 4\% within RFG approach and an increase of about 2\% using HO and NO spectral functions, almost independent of the neutrino energy.
\item[3.] All the different approaches considered in this work, based on the impulse approximation, underpredict the MiniBooNE data for the flux-averaged CCQE $\nu_\mu$($\overline{\nu}_\mu$)$+ ^{12}$C differential cross section per nucleon and the total cross sections, although the shape of the cross sections is represented by the NO+FSI, HO+FSI and RFG+FSI approaches. We note that the comparison of our results for the total cross section with the MiniBooNE data in the case of antineutrino scattering shows much better agreement than in the neutrino case.

    Here we emphasize that all models used give results that are compatible with the NOMAD data. This result points to the importance of a careful evaluation of non-impulsive contributions, like the ones associated to meson-exchange-currents, and of their evolution with energy.

\end{itemize}

Along this line, we should comment on the general problem that none of the models explored in the present work agrees with the MiniBooNE data and not even SuSA does. This could be due to important ingredients that are missing in the considered theoretical models and would improve the agreement with the MiniBoonE data. In our opinion, the $2p-2h$ MEC contribution may be responsible for the present discrepancy, in agreement with the results of Refs.~\cite{Amaro2011151, PhysRevLett.108.152501, PhysRevC.81.045502, Nieves201272}. This is corroborated by the fact that a similar disagreement with theory appears when the phenomenological scaling function SuSA is used. The latter is a purely longitudinal quasielastic response, and $2p-2h$ MEC contributions should not contribute to it when properly extracted from quasielastic electron scattering data, but could contribute to quasielastic neutrino-nucleus scattering because of the axial current.

\section*{Acknowledgements}

This work was partially supported by Spanish DGI and FEDER funds (FIS2011-28738-C02-01, FPA2010-17142), by the Junta de Andalucia, by the Spanish Consolider-Ingenio 2000 program CPAN (CSD2007-00042), by the Campus of Excellence International (CEI) of Moncloa project (Madrid) and Andalucia Tech, by the Istituto Nazionale di Fisica Nucleare under Contract MB31, by the INFN-MICINN collaboration agreement (AIC-D-2011-0704), as well as by the Bulgarian National Science Fund under contracts No. DO-02-285 and DID-02/16-17.12.2009. M.V.I. is grateful for the warm hospitality given by the UCM and for financial support during his stay there from the SiNuRSE action within the ENSAR european project. A.N.A. acknowledges financial support from the Universidad de Sevilla under the Program: ``IV Plan Propio de Investigaci\'on. Movilidad de Investigadores''.

\bibliography{rif}
\end{document}